# A Time-Triggered Communication Method Based on Urgency-Based Scheduler in Time-Sensitive Networking

**Feng Luo[a,*], Yunpeng Li[a], Zitong Wang[a], Yi Ren[a], Yingpeng Tong[a], Zhouping Zhang[b], and Qin Liu[c]**

[a] School of Automotive Studies, Tongji University, Shanghai, China

[b] Nanchang Automotive Institute of Intelligence and New Energy, Tongji University (NAIT), Nanchang, China

[c] Jiangxi Key Laboratory of Intelligent Connected Vehicle and Powertrain System, Nanchang, China

## Abstract

The development of the automotive industry and automation has led to a growing demand for time-critical systems to have low latency and jitter for critical traffic. To address this issue, the IEEE 802.1 Time-Sensitive Networking (TSN) task group proposed the Time-Aware Shaper (TAS) to implement Time-Triggered (TT) communication, enabling deterministic transmission by assigning specific time windows to each stream. While Fixed Routing and Waiting-Allowed (FR-WA) scheduling algorithms offer flexibility, they suffer from inefficiencies in solution time and scalability. This study analyzes TAS implementation challenges, emphasizing how network scale expansion increases computational constraints. We propose an Urgency-Based Scheduler method (TT-UBS) to address these limitations to enhance deterministic transmission and computational efficiency under anomalies. A novel scheduling algorithm for TT-UBS parameter determination is developed, alongside simulations and comparative evaluations. Results show that TT-UBS guarantees deterministic traffic delivery while reducing solution time by 98.22% in test scenarios compared to traditional approaches. The methodology is extended to other scheduling algorithms to assess efficiency improvements. This advancement supports TSN's application in mission-critical systems by optimizing time-triggered communication performance and enabling reliable network deployment. The framework demonstrates significant potential for real-time in-vehicle networks requiring latency and jitter control.

## 1 Introduction

The automotive and automation industries are experiencing a surge in the proliferation of time-critical systems, necessitating precise timing, efficient communication, and low latency to ensure the reliable operation of advanced functionalities such as autonomous driving and real-time control. To achieve efficient communication and reliable data transmission in time-critical systems, traditional communication protocols are no longer sufficient to meet the demand, which requires low latency and jitter in transmitting critical traffic [1]. Therefore, Time-Triggered (TT) communication is proposed to guarantee deterministic traffic transmission. TT communication is a time-based method that provides high determinism and predictability for traffic transmission by providing a specific transmission schedule for messages in the network to prevent message collisions.

The IEEE 802.1 Time-Sensitive Networking (TSN) task group proposes a set of standards to provide high reliability and certainty for the transmission of traffic in the network [2]. These standards relate to time synchronization, bounded low latency, high reliability, and network management, providing network designers many options. Among these standards, IEEE 802.1Qbv-2015 (now enrolled in the IEEE 802.1Q-2018 standard [3]) introduces the concept of Time-Aware Shaping (TAS) to realize TT communication. TAS controls the opening and closing of each queue's gate through the Gate Control List (GCL) when the device clocks in the network are synchronized. Therefore, the streams in the queues can be transmitted in the time windows defined by the GCL.

One research direction for TT communication in TSN is End-to-End (E2E) latency analysis. The work in [4] performs a schedulability analysis of TAS using Compositional Performance Analysis (CPA) to derive its worst-case latency bounds. Some studies compare TAS with Asynchronous Traffic Shaper (ATS) and the Cyclic Queue and Forwarding (CQF) mechanism (proposed in IEEE 802.1Qch) to derive their advantages and the applicable scenarios [5, 6]. Then, the work in [7] considers combining the Credit-Based Shaper (CBS) and TAS and computes the worst-case latency of classes A and B streams. The author in [8] also analyzed various scenarios of using TAS and frame preemption together. The authors in [9] analyzed the worst-case response time for using CQF and the preemption mechanism and validated the analysis results through simulation in [10], deriving recommendations for parameter configuration. Moreover, the work in [11] proposes a TSN partitioning system using TAS that allows the introduction of new traffic in the in-vehicle network without affecting the existing traffic.

Although the mechanism of TAS itself is simple, there are difficulties in getting the proper GCL to deploy the TT communication on the network. It can prove to be a Non-deterministic Polynomial-time (NP)-hard problem [12]. Therefore, another research direction for TT communication in TSN is deriving a suitable time window design solution. Scheduling algorithms can be classified into two categories by whether or not they incorporate routing solutions: Fixed-Routing (FR) algorithms and Joint-Scheduling (JS) algorithms [13]. Unlike [13], we collectively refer to the scheduling algorithms of joint tasks and routes as JS algorithms. The FR algorithm treats routes as known conditions (one of the inputs to the solution), focusing on scheduling-related parameters. In contrast, the JRS algorithm considers the scheduling algorithm's solution in conjunction with other solutions. For example, the work in [14] proposed a learning-based scalable scheduling and routing co-design (LSSR) architecture for TSN. Yang *et al*. [15] proposed a scheduling algorithm for joint task and traffic scheduling. Moreover, scheduling algorithms can also be classified into No-Waiting algorithms (NW) and Waiting-Allowed (WA) algorithms by whether or not they allow frames to wait in the queues [13]. The NW scheduling algorithm requires that all frames cannot wait in a queue during forwarding (no queue delay during frame transmission), thus reducing the E2E latency of frame transmission. Durr *et al.* [16] proposed a scheduling algorithm using an Integer Linear Programming (ILP) solver that supports the no-wait transmission mode to reduce TT traffic's E2E latency while increasing the bandwidth for BE traffic. Zhang *et al.* [17] proposed a method based on divisibility theory to characterize the conflicts and dependencies of streams for parameter solving of CQF mechanisms. Subsequently, they applied the technique to parameter solving for TAS in NW mode (NW-TAS) [18]. They considered the virtual queue architecture proposed in [19] in their implementation to achieve efficient parameter solving. The WA scheduling algorithm allows frames to wait in the queue for some time, resulting in a more extensive solution space. The most representative study is the work in [20], where they constructed the primary constraints under the WA-TAS model, with the most notable contribution being the proposed the frame isolation constraints. Subsequently, many researchers solve scheduling-related parameters under the WA-TAS model, encompassing algorithms based on solvers [21, 22] and heuristic methods [23-25]. Although NW scheduling algorithms can provide for the reduction of the E2E latency of frames, they are limited to a smaller space of solutions compared to WA scheduling algorithms, which results in WA scheduling algorithms being more flexible and able to be applied to a broader range of scenarios (with schedulability advantages) [13]. Therefore, in this paper, we focus on WA scheduling algorithms for a more expansive solution space. Moreover, compared to FR scheduling algorithms, JS algorithms typically result in lower schedulability ratios due to computational overhead. Therefore, we focus on FR-WA scheduling algorithms in this paper.

* Corresponding author..
E-mail address: luo_feng @tongji.edu.cn
Postal address: NO.4800, Cao'an Highway, Jiading District 201800, Shanghai, China.

In real-time systems, although TAS can theoretically provide deterministic latency, this determinism may not be achievable due to anomalies caused by physical factors and external attacks, such as traffic loss or latency anomalies, which contradicts the goal of TSN. In the case of traffic anomalies, under the TAS mechanism, delayed frames may occupy the transmission window of subsequent frames by missing the expected transmission window (discussed in the following), which may lead to an increase in E2E latency for the following traffic. Although some of the studies mentioned above have analyzed E2E latency for TAS, they have yet to solve the problems faced by TAS in anomalous communication situations. Moreover, even though FR-WA scheduling algorithms have a more comprehensive solution space, their efficiency still needs improvement. Yun *et al.* [19] eliminated the enqueue conflicts by proposing the Virtual-Queue Switching (VQS) architecture with Parallel Shared Memory (PSM) for switches. Similarly, in the process of parameter solving for the TT communication mechanism proposed in this paper, the solution's efficiency can also be improved by eliminating the frame isolation constraints. However, the work in [19] did not consider the traffic anomaly scenario. Furthermore, our UBS-based improvement can provide a more standard, generic architecture and straightforward implementation. To the best of our knowledge, this paper is the first to propose the use of a UBS architecture in TT communication combined with the PSFP mechanism to mitigate the negative impact of traffic anomalies and improve the efficiency of parameter solving.

The main contributions of this study are as follows.

- **TT Communication Method Based on UBS Architecture:** Proposes a UBS-based TT communication method (TT-UBS) that adapts asynchronous traffic scheduling mechanisms (UBS) to synchronous transmission scenarios. This approach integrates IEEE 802.1Qci PSFP protocols with UBS architecture to classify traffic flows into shaped queues, enabling per-stream deterministic scheduling through queue-level time window control.
- **Lookup Table-Based Frame Timing Control Mechanism:** Introduces a novel shaper module within the UBS framework that dynamically manages frame issuance timing using a lookup table driven by precomputed parameters. This mechanism employs queue length threshold checks to isolate anomalous frames, ensuring reliable transmission of subsequent traffic.
- **End-to-End Latency Analysis Framework:** Establishes a formal mathematical model of E2E latency in TT-UBS transmission paths, conducts component-level formal verification (shapers, schedulers, switches), and quantifies the impact of critical configuration parameters on latency performance.
- **Lightweight Scheduling Algorithm Optimization (SMT-WA-NFIC):** Proposes an enhanced SMT-WA-NFIC algorithm that removes frame isolation constraints for efficient parameter computation. Extends the methodology to AT/LSTB algorithms through simulations, demonstrating 98.22% improvement in solution time compared to conventional approaches across standardized test scenarios.

The remainder of this paper is outlined below. The second part gives an overview of TAS and summarizes the frame isolation constraints in the FR-WA scheduling algorithms. Additionally, an analysis is conducted to investigate how the number of constraints changes as the network scale increases and potential problems of TAS in the presence of traffic anomalies are identified. The third part proposes a TT communication method based on the UBS mechanism to avoid these problems and increase the solver speed. A shaper based on a look-up table has been proposed to allocate appropriate transmission times for frames and discard those that have exceeded their time limits. A formal analysis of frame latency and performance analysis of the TT-UBS mechanism is also provided. We also present a scheduling algorithm to derive the parameters of the proposed mechanism. In the fourth part, the mechanism is evaluated using the OMNeT++ simulator under traffic anomaly scenarios, and the solution speed is compared with others.

## 2 Key Mechanisms and Problem Analysis

### *2.1 TAS in IEEE 802.1Qbv*

In the IEEE 802.1Qbv standard, gates are assigned to each priority queue and control the transmission of packets in the queue. The open and closed states of the gates are determined by the GCL, which is cyclically repeated with a fixed cycle time. With this gate control mechanism, the TAS mechanism can allocate specific time windows for transmitting different streams in the network.

Figure 1 shows that the GCL opens the gates for Queue 0 and Queue 7 in separate time windows, enabling packets to be transmitted in their respective queues. During the transmission process, the gates of other queues remain closed to prevent interference from packets in those queues. This approach ensures deterministic stream transmission while maintaining fairness and avoiding stream collisions.

As described above, the TAS mechanism is not overly complex, as it uses the GCL to regulate the transmission of packets in the queues. Therefore, the critical aspect of deploying this mechanism is configuring an appropriate GCL for different network scenarios to ensure traffic transmission meets the requirements. This can be achieved through scheduling algorithms, and as previously mentioned, this paper focuses primarily on FR-WA scheduling algorithms.

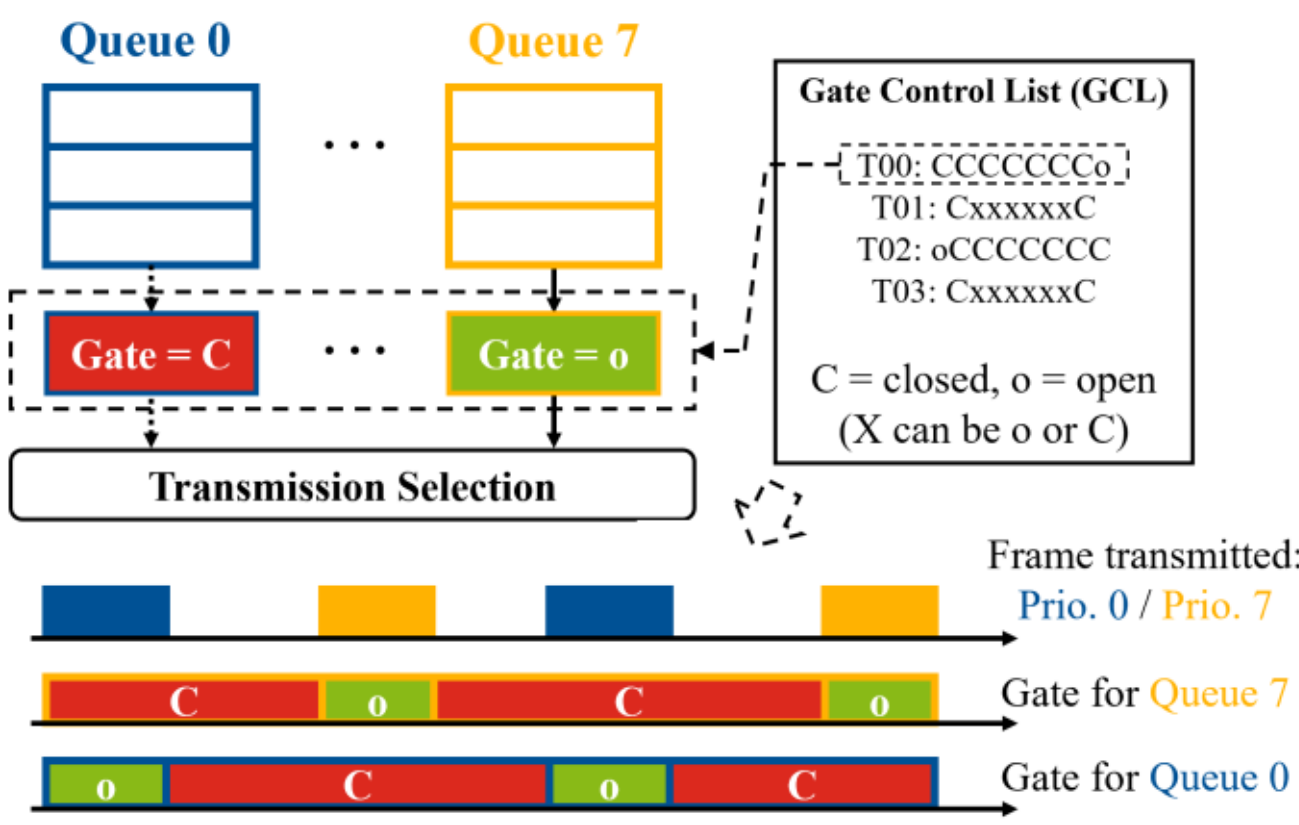

Figure 1. Illustration of the working principle of the TAS mechanism proposed in IEEE 802.1Qbv.

## *2.2 Enqueue Restrictions in Scheduling Algorithms*

### 2.2.1 General Definition

We abstract the network as a directed graph $G(V,E)$, where $V$ is a set of vertices indicating the network's end stations and switches. $E$ is a set of directed edges between two vertices. Since Ethernet links are full-duplex, the physical link between nodes $v_a$ and $v_b$ can be expressed as $[v_a,v_b]$ and $[v_b,v_a]$, where the former node is the source node and the latter is the destination node. $[v_a,v_b]$ and $[v_b,v_a]$ belong to $E$, with properties defined by the tuple $\langle s,d,p,c\rangle$, where $s$ is the transmission speed, $d$ is the transmission delay on the medium, $p$ is the process delay of the source node, and $c$ is the number of available priorities on the link. Here we use the "." symbol to indicate getting an element of a tuple. For example, $[v_a,v_b].s$ indicates the transmission speed of the link $[v_a,v_b]$.

A stream can be defined by the tuple $\langle T,L,q,e2e,jitter\rangle$, denoting the period, the length in bytes, the transmission queue, the maximum allowed E2E latency and the allowed jitter of the stream. We define the set of streams transmitted on the link $[v_a,v_b]$ as $S^{[v_a,v_b]}$ and the hyper-period $hp^{[v_a,v_b]}$ of these streams can be denoted as Equation (1).

$$hp^{[v_a,v_b]} = lcm\left\{s_i.T, s_i \in S^{[v_a,v_b]}\right\} \tag{1}$$

The "lcm" in the formula indicates the operation of taking the least common multiple. Appropriate GCL can ensure the network's orderly and conflict-free transmission of traffic. For the GCL, it is necessary to construct several constraints, such as the frame constraints, the link constraints, the flow transmission constraints, the E2E constraints, and the frame isolation constraints in [20]. It is important to note here that in frame-based mode, the jitters of frames are bounded by the frame isolation constraints [20].

In a hyper-period on the link $[v_a,v_b]$, the set of frames of stream i can be denoted as $F_i^{[v_a,v_b]}$. The jth frame of the stream i transmitted on the link $[v_a,v_b]$ can be expressed as $f_{i,j}^{[v_a,v_b]}$ with the properties defined as the tuple $\langle \phi,T,L\rangle$, denoting the offset, the period, and the transmission duration of the frame.

### 2.2.2 The Frame Isolation Constraint

The frame isolation constraints specify that only one frame can be stored in one queue to prevent uncertain transmissions [20], as shown in Figure 2. It can also be denoted as Equation (2). In the scenario depicted on the right side of Figure 2, the second frame of Stream 1 is still in the queue when the second frame of Stream 2 arrives, resulting in two frames simultaneously being in the same queue. In that case, the time they enter the queue is not deterministic due to factors such as uncertainty in the device transmission delay. These constraints also ensure that if frames that arrive simultaneously are stored in different queues, this requirement must not be satisfied.

$$\begin{aligned}
&\forall [v_a,v_b] \in E, \forall s_i, s_j \in S^{[v_a,v_b]}, i \neq j \\
&\forall f_{i,k}^{[v_a,v_b]} \in F_i^{[v_a,v_b]}, f_{j,l}^{[v_a,v_b]} \in F_j^{[v_a,v_b]} \\
&\forall \alpha \in \left[0, hp^{[v_a,v_b]}/s_i.T - 1\right], \forall \beta \in \left[0, hp^{[v_a,v_b]}/s_j.T - 1\right] \\
&\left( f_{j,l}^{[v_a,v_b]}.\phi + \beta \cdot f_{j,l}^{[v_a,v_b]}.T + \delta \leq \right. \\
&\left. f_{i,k}^{[v_x,v_b]}.\phi + \alpha \cdot f_{i,k}^{[v_x,v_b]}.T + [v_x,v_b].d \right) \vee \\
&\left( f_{i,k}^{[v_a,v_b]}.\phi + \alpha \cdot f_{i,k}^{[v_a,v_b]}.T + \delta \leq \right. \\
&\left. f_{j,l}^{[v_y,v_b]}.\phi + \alpha \cdot f_{j,l}^{[v_y,v_b]}.T + [v_y,v_b].d \right) \vee \\
&\left( s_i^{[v_a,v_b]}.p \neq s_j^{[v_a,v_b]}.p \right)
\end{aligned} \tag{2}$$

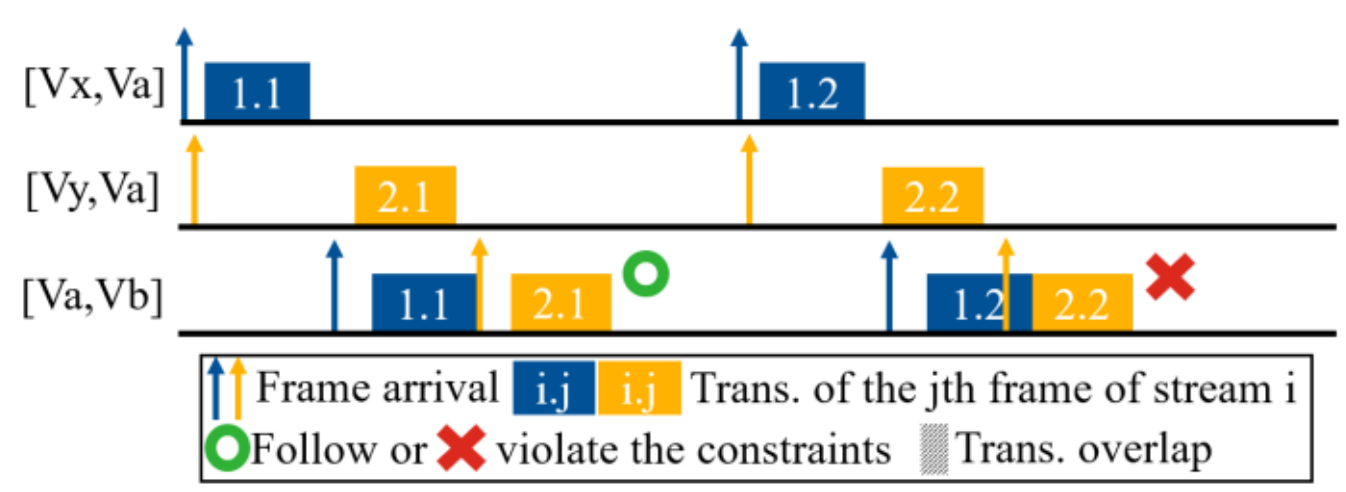

Figure 2. Illustration of the frame isolation constraints (assuming that the two streams are planned into the same queue on the link [va,vb]).

In the frame-based approach, each transmitted frame has a corresponding time window allocated to it. If a frame misses its allocated time window (i.e., the time window has already closed by the time the frame arrives at the port), it will affect the transmission of subsequent frames, and in some cases, it may not be transmitted at all. Moreover, in the transmission of frame-based approach, the jitter of a frame is deterministic (or predictable), which aligns with the requirement of predictability in hard real-time systems. The frame isolation constraints are proposed in [20] to specify the offsets at which upstream devices send frames to ensure the deterministic transmission order of frames at the current port. Moreover, to our knowledge, all existing FR-WA scheduling algorithm that ensure the order of frame enqueue rely on the frame isolation constraints. We are the first to propose alternative methods for guaranteeing the order of frame transmission. For one of the focuses of this paper, we provide a more detailed description of the significance of the frame isolation constraints by presenting an example, as shown in Figure 3. In the scenarios depicted in Figure 3, packets from two different links [vx,va] and [vy,va] need to be sent on the same link. We need the order of the transmitted packets to be fixed to ensure that there is no jitter during the packet transmission process.

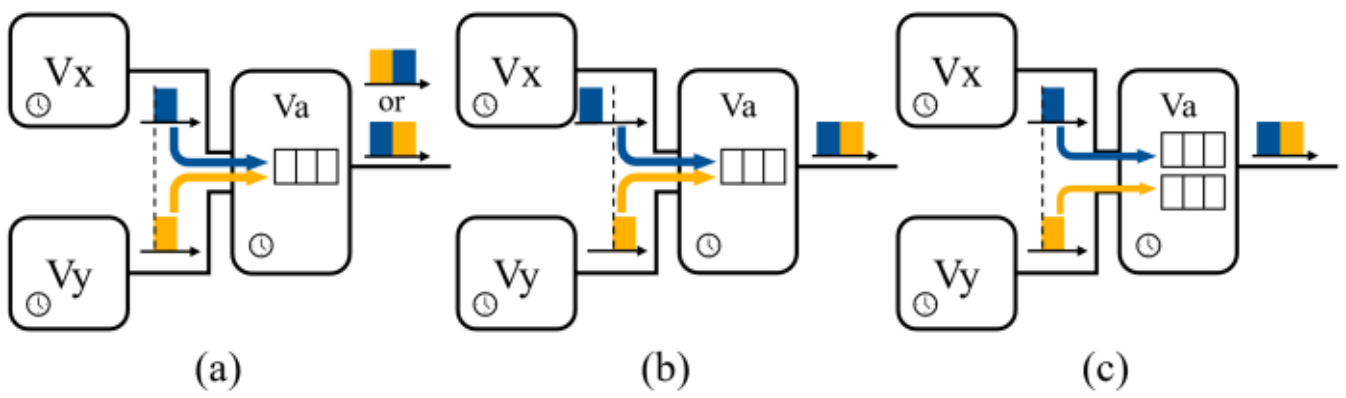

Figure 3 The transmission results for three different scenarios: (a) when two packets arrive simultaneously and enter the same queue, (b) when one packet arrives after another packet has been transmitted and both packets enter the same queue, and (c) when two packets arrive simultaneously and enter different queues.

As illustrated in Figure 3(a), if these two packets arrive simultaneously and are stored in the same queue, the order in which they are transmitted will not be guaranteed because the switch's internal forwarding delay is usually variable. Moreover, packet transmission may exhibit jitter when packet loss occurs in the network. If the queue has permission to hold two packets simultaneously, losing the first packet can cause the second packet to be transmitted earlier, potentially resulting in additional jitter.

In Figure 3(b), two packets from different links are stored in the same queue but do not arrive simultaneously. Assuming that in this scenario, the second packet arrived after the first packet had been transmitted, unlike the scenario depicted in Figure 3(a), the order in which the two packets are transmitted is deterministic. Even if the first packet is lost, there is no jitter in the transmission of the second packet. Combining Figures 3(a) and (b), we can conclude that this is why the frame isolation constraints require that two packets cannot simultaneously exist in the queue, as it guarantees that there is no jitter in the packet transmission process.

In Figure 3(c), packets from different links are stored in different priority queues within the switch. Even if these packets arrive simultaneously, they can still be transmitted in a deterministic order because the GCL controls when each queue can send packets. Therefore, the frame isolation constraint specifies that if two packets are assigned to different queues, they do not need to meet the requirement of arrival time. In such a scenario, the GCL can entirely control the order in which the two packets are transmitted, thereby avoiding any uncertainty in the packet transmission.

Based on the analysis of Figure 3, we can conclude that the frame isolation constraints require the arrival times of two packets stored in the same priority queue to satisfy the condition that only one packet can exist in one queue at a time. Otherwise, the order of packet transmission may become uncertain due to the impact of the switch's internal forwarding delay. When two packets are stored in different priority queues, their arrival times do not need to meet any requirements. It is evident that when multiple packets from different sources share the same link, the frame isolation constraint needs to be added to the scheduling algorithm. As the number of devices and traffic increases, the number of such constraints will also increase. In the next section, we will show how the number of frame isolation constraints changes as the network expands.

## *2.3 Analysis of Existing Problems in Deploying TAS*

### 2.3.1 Problems under Traffic Anomaly

Although TAS can provide deterministic E2E latency in an ideal case, some abnormal traffic transmissions in real-time systems, such as traffic loss or traffic experiencing abnormal latency, are difficult to avoid. The abnormal transmission of certain traffic can jeopardize the transmission of other traffic in the network, leading to a loss of determinism in network communication. This contradicts the objective of TSN.

The abnormal traffic transmission can be due to the failures of devices in the network. When hardware or software failures occur in network devices, it can result in the loss and delay of transmitted frames within the network. Moreover, wired networks rely on physical connections, and therefore, poor connections, cable damage, or electromagnetic interference during frame transmission may also contribute to frame loss. Additionally, high network loads exceeding the processing capacity of network devices can lead to abnormal traffic transmission.

In addition to the aforementioned network device failures, the security threats faced by networks should not be overlooked. The work in [26] utilized the STRIDE model to analyze the potential threats faced by various protocols in TSN. It pointed out that packets within the network may be susceptible to delaying attacks, particularly those related to time synchronization. Studies on the presence of time delay attacks in networks were also conducted in [27, 28]. Some attackers may also impose delays on frames in the network through the Man-In-Middle (MIM) attack. The work in [27] indicated that an external entity with expertise and appropriate equipment or a malicious individual with access to the network can choose suitable attack locations (such as links in the vehicle backbone network) to execute Time Delay Attacks (TDA) and avoid detection. Furthermore, the work in [29] highlights that the scheduling mechanisms in TSN may face risks of tampering or forging configuration packets, which can result in abnormal traffic transmission within the network.

TAS may cause streams to be transmitted without meeting the latency requirement in the case of traffic anomaly, as shown in Figure 4.

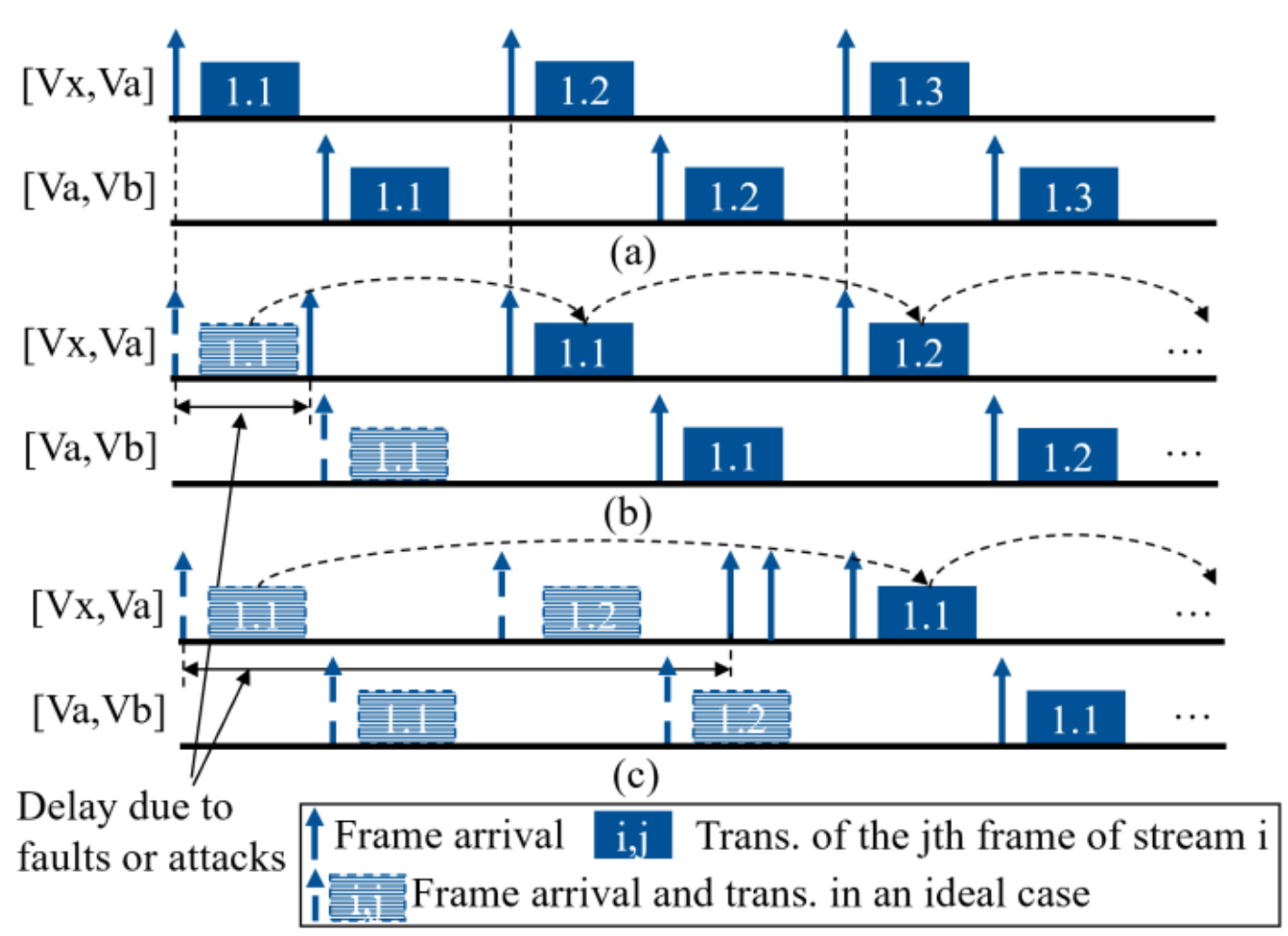

Figure 4. Frame transmission (a) under an ideal case, (b) under a traffic anomaly case, and (c) under a traffic anomaly case with a long delay time.

Figure 4(a) depicts a scenario where a packet is transmitted from the link [vx,va] to the link [va,vb] without any issues. In this scenario, each packet is transmitted according to a pre-planned schedule based on the GCL.

Figure 4(b) and (c) shows that when the first frame arrives late at the link [Vx,Vb] due to some condition (assuming it has missed its time window), it can only wait for the next window to open in the queue, which leads to the blocking of frames in the queue and does not meet the E2E latency requirements. Especially in Figure 4(c), as the delay time increases, the impact on frame transmission becomes more significant. Since the Per Stream Filtering and Policing (PSFP) mechanism proposed in IEEE 802.1Qci can operate on abnormal streams, we consider combining it into the proposed method.

### 2.3.2 Increasing Number of Constraints

Since most parameter solutions are obtained as feasible solutions within the limits of the constraints satisfied, the number of constraints affects the solution's efficiency. The increase in network size is usually reflected in the increase in network devices and transmission traffic. Based on the analysis of the constraints mentioned above, it can be determined that, given a fixed network topology and traffic, the number of constraints to be added can be determined. For instance, when a stream is transmitted from one link to a downstream link, a certain number of constraints, should be added for ensuring that the transmission offset of every frame of this stream in a hyper-period satisfies the Link Constraint. To derive the variation of the number of each constraint with the increase of network devices and traffic, we counted the number of each constraint in chained topologies using the randomly generated traffic.

In the chained topologies, where each switch connects three end stations, we increase the number of switches in the network from 1 to 10 to build the case of increasing network devices, as shown in Figure 5. Then, we increase the number of streams from 5 to 95 in steps of 5 in each topology. In this way, we can obtain 190 network scenarios (with a fixed number of network devices and streams). For each network scenario, we calculate the number of constraints (including the Frame Constraint, the Link Constraint, the Flow Transmission Constraint, the E2E Constraint, and the Frame Isolation Constraint) by traversing the links and streams in the network. Moreover, since the streams were generated randomly (their size, period, and routing are randomly generated), certain attributes of the traffic, such as the period and routing, may have an impact on the number of constraints. Therefore, we repeated the process of counting the number of constraints 500 times for each network scenario. We calculated the average of the constraint numbers under each case to make the results more generalizable, as shown in Figure 6. The size of the streams is chosen randomly from the set [400 Bytes, 600 Bytes, 800 Bytes, 1000 Bytes, 1500 Bytes], and the period is selected from the set [10 ms, 20 ms]. Note that we do not represent the variation of all constraints in Figure 6 to express more clearly the factors affecting the number of constraints.

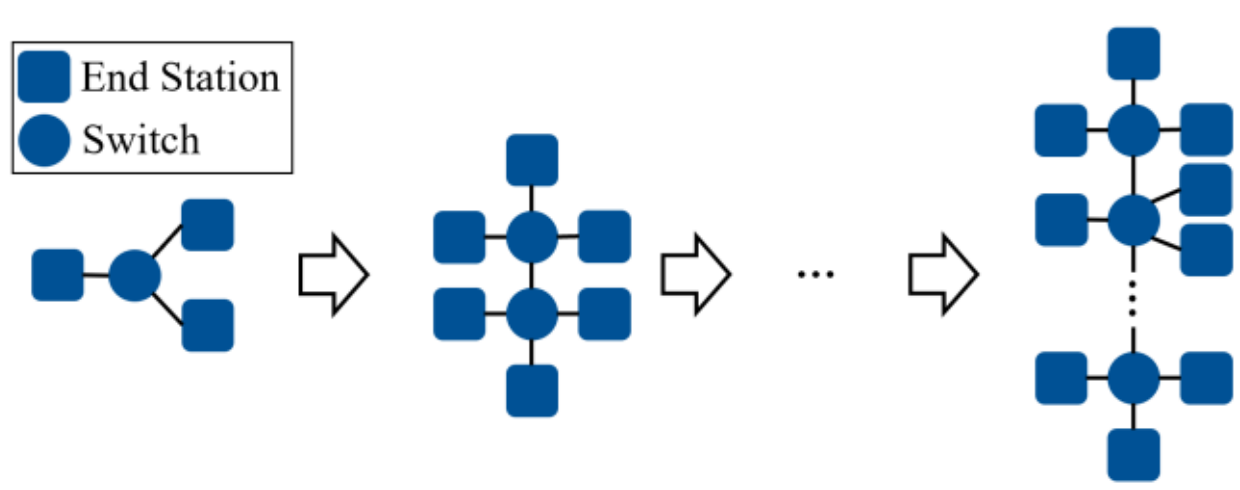

Figure 5. Increasing the number of switches in the network topology where each switch connects three end stations.

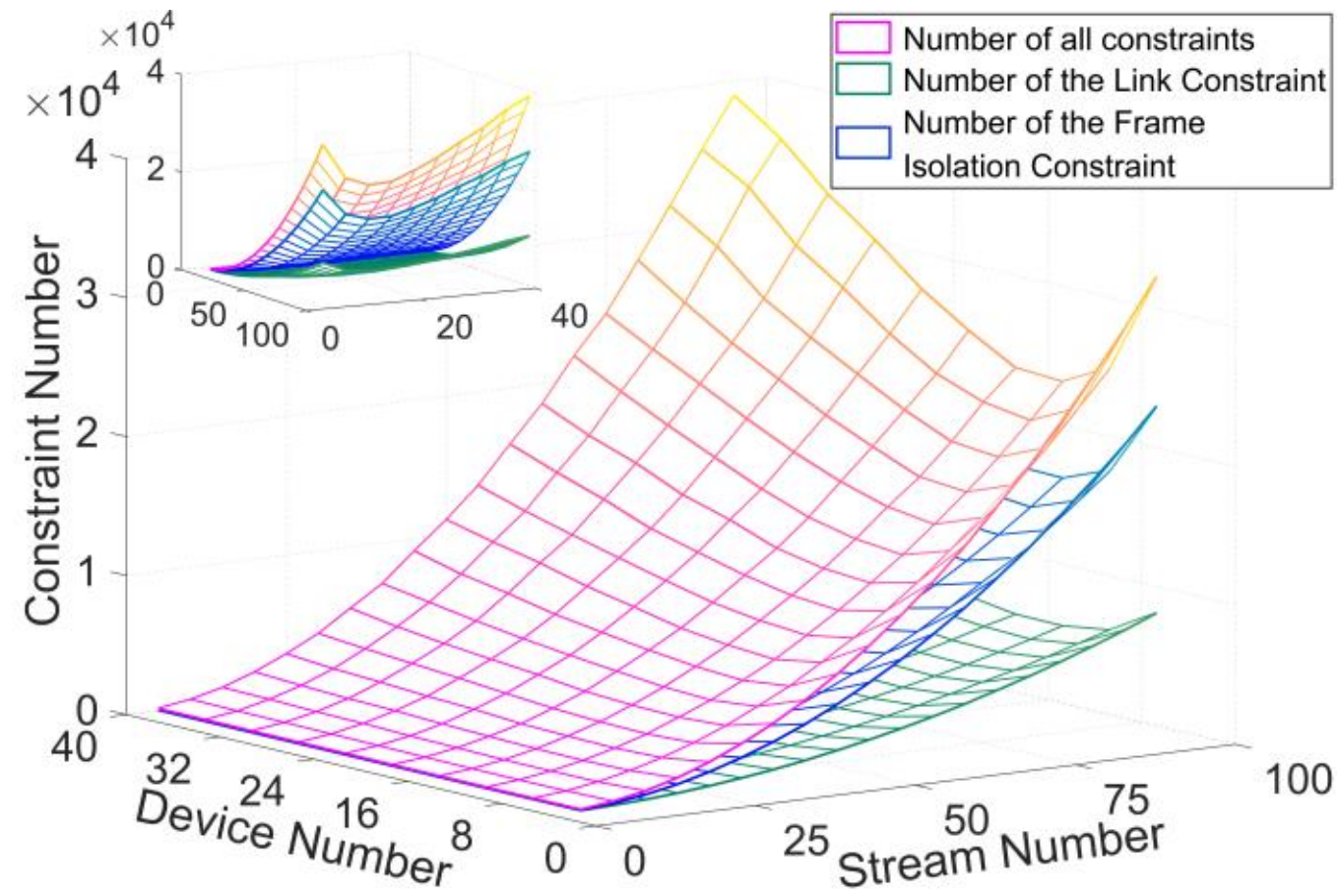

Figure 6. Number variation of all constraints, the link Constraints, and the frame isolation constraint with increased stream number and device number in the network.

As seen in Figure 6, the constraint number increases significantly with the stream number growth. It can be seen that when the number of devices is four and the number of streams is 5, the total number of constraints is in the tens. However, when the number of devices is 40, and the number of streams is 100, the total number of constraints can be around 40,000.

Furthermore, as shown in Figure 6, when the number of streams is fixed, the total number of constraints initially decreases and then increases as the number of devices increases. This is because when the device number is too small, more streams are transmitted on one link, increasing the constraint number. When the device number increases, streams transmitted on each link may decrease, leading to fewer constraints. However, when the device number is too big, the transmission hops of the streams grow accordingly, which can also increase the constraint number.

Additionally, Figure 6 illustrates that the number of constraints is primarily affected by the link constraints and the frame isolation constraints, with the latter having a more significant impact. Therefore, simplifying the frame isolation constraints may effectively reduce the solution time, making it one of the key goals of this paper.

## 3 The Proposed TT-UBS Mechanism

### *3.1 UBS Model Architecture*

UBS was initially designed to guarantee low latency in switched Ethernet under an asynchronous case (clocks of devices in the network are not synchronized) while keeping implementation complexity low [30].

UBS has a hierarchical queue architecture, including shaped and shared queues, as shown in Figure 7. The shaped queues follow the First-In-First-Out (FIFO) principle. They are assigned with fixed priority according to the upstream sources. The shared queues are also assigned with scheduler internal fixed priority following the FIFO principle. Firstly, streams from different input ports are passed through the stream filters and gates to enable differentiation between different streams. Streams of the same priority in the transmitters and receivers from the same port will be assigned to the same shaped queue. The streams from the shaped queues are mixed in the shared queue according to their priorities in the device after passing the shapers. The current scheduling algorithms used in the shapers of UBS are Length-Rate Quotient (LRQ) and Token Bucket Emulation (TBE). Both limit streams' transmission rates by the token bucket-like approach to implementing ATS. Finally, the transmission selection module sends the streams in the shared queues based on the strict priority algorithm. Unlike other shaping mechanisms in TSN, UBS is performed on individual streams rather than priorities, which gives it more configuration space for more complex use cases.

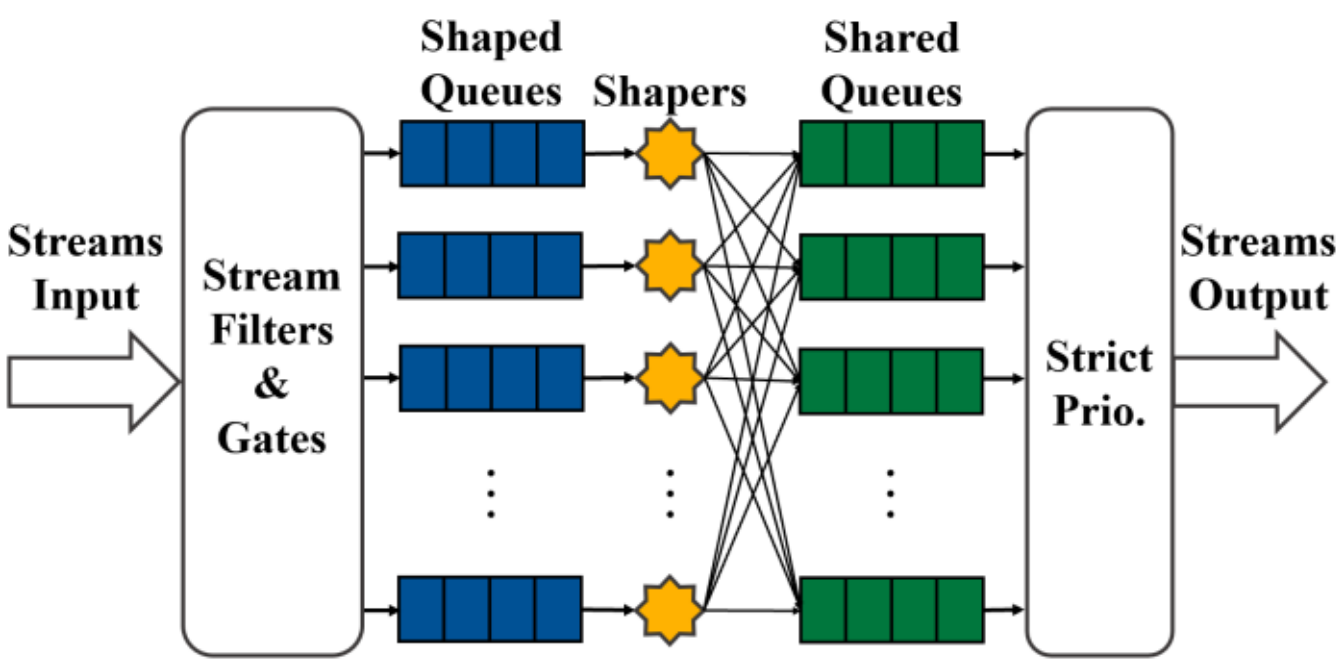

Figure 7. Illustration of the transmission model in UBS.

### *3.2 TT Communication Method based on UBS*

Based on the analysis above, it is clear that the current TAS mechanism experiences reduced solution efficiency when using constraint-based scheduling algorithms to solve deployment plans due to the impact of frame isolation constraints. Furthermore, in situations of abnormal traffic flow, TAS can also affect the QoS of traffic transmission in the network. Therefore, this paper introduces a new shaper to implement TT communication based on the UBS architecture, effectively reducing the solution time when the deployment scheme is derived using constraint-based scheduling algorithms. It can also ensure the QoS of traffic in the network is not affected in the case of traffic anomaly.

Figure 8 illustrates the process that a frame undergoes after entering the TT communication mechanism proposed in this paper. The proposed shaper differs from the traditional LRQ and TBE (token bucket-based shaper) used in conventional UBS. It controls the transmission of frames in its corresponding Shaped Queue through a pre-defined table called the *ShaperOffsetTable*. This table stores the offset times within the hyper-period for each frame's transmission. The functionality of the shaper proposed in this paper is reflected in the "Sub-processes" section in Figure 8. It should be noted that although UBS is designed to implement asynchronous traffic shaping, we use it in clock synchronization scenarios to implement TT communication in this paper. The shared queues mentioned in the TT-UBS mechanism and the priority queues in TAS refer to the same entity. The detailed description of the process illustrated in Figure 8 is as follows:

- When a frame arrives at a port, it is first passed to the Stream Filter O to match its attributes, such as MAC address and priority, against those defined in the stream_handle of the Stream Filter O. If the frame's attributes match those defined in the stream_handle, it will be passed to the corresponding Stream Gate P, as shown in Figure 8. Otherwise, it will be passed to the next Stream Filter and checked until matched or discarded.
- When the frame arrives at Stream Gate P, the gate's open or closed state determines whether the frame is discarded or passed through. In this mechanism, we do not want the frame discarded at the gate, so we set the gate to be in an open state by default. Therefore, the frame will be passed to Shaped Queue Q instead of discarded.
- After passing through Stream Filter O and Stream Gate P, the frame can be stored in Shaped Queue Q, as shown in Figure 8. As soon as the frame is stored in Shaped Queue Q, the related Shaper R starts working, as indicated in the Sub-Processes of Figure 8.
- The abovementioned process ensures that different streams are stored in separate shaped queues, meaning each shaped queue exclusively holds frames from a single stream. The frame transmission is periodic. To meet the requirements of the frame constraint and E2E latency, the shaped queue must send out the frames within its cycle time. Consequently, the scenario where two or more frames coexist simultaneously in the same shaped queue is impossible. In the proposed mechanism, frames will be transmitted at the appropriate time. However, frames that miss the appropriate transmission time will remain in the queue. In such a case, it would be considered that the previous frames in the queue Q are delayed (as they missed the appropriate transmission time), leading to the discarding of previous frames in the queue Q to retain the most recently arrived frame. Therefore, in the Sub-Processes section of Shaper R, an assessment is made to determine whether there are two or more frames in queue Q. If this condition is met, the previous frames are discarded. The parameter named *CurrentGlobalTime*, which represents the current global time in the network after synchronization, is obtained as soon as the frame is stored in Shaped Queue Q. The shaper R can calculate the current offset of the frame within its corresponding cycle time (*CurrentOffset*), based on the hyper-period that encompasses all the streams on the output port (*CycleTime*). Subsequently, the Shaper R retrieves the *EligibilityOffset* for the frame in the current cycle by searching the *ShaperOffsetTable*. *EligibilityOffset* represents the appropriate sending offset time for the frame and is then compared to the previously calculated *CurrentOffset*. If the frame's *EligibilityOffset* is less than the CurrentOffset, the frame is deemed a timeout frame, which could potentially impact the sending of subsequent messages, as depicted in Figure 4(c). To mitigate this issue, the Shaper R discards the frame. On the other hand, if the *EligibilityOffset* for the frame is greater than the *CurrentOffset*, the shaper will wait until the *EligibilityTime*, which is derived from mapping the *EligibilityOffset* to the global time, before sending the frame. This ensures that the frame is sent at the optimal time determined by the scheduling algorithms, thereby avoiding network congestion.
- When the frame arrives at the Shared Queue S, unlike in the TAS mechanism, the gate of the Shared Queue S that stores the frame is always open to ensure that the frame is sent out at the expected time.

It is important to note that, unlike UBS, the TT-UBS mechanism includes a gate after each shared queue, which operates similarly to TAS. Moreover, the gate for the shared queue, through which the time-critical frames pass, is constantly kept open to ensure the transmission of frames following the *EligibilityOffset* of the corresponding shaped queue. The above process is primarily designed for time-critical frames. As shaped queues can store these frames, the number of shared queues used to keep them can be minimized to the smallest possible number, which is one. This also provides more shared queue choices for other types of traffic, such as best-effort traffic, while reducing the allocation of shared queue resources. For best-effort traffic, a Stream Filter can be set up to transfer it to the corresponding shared queue directly. The gating design of these queues is similar to TAS, and they only need to be closed when time-critical frames are being sent.

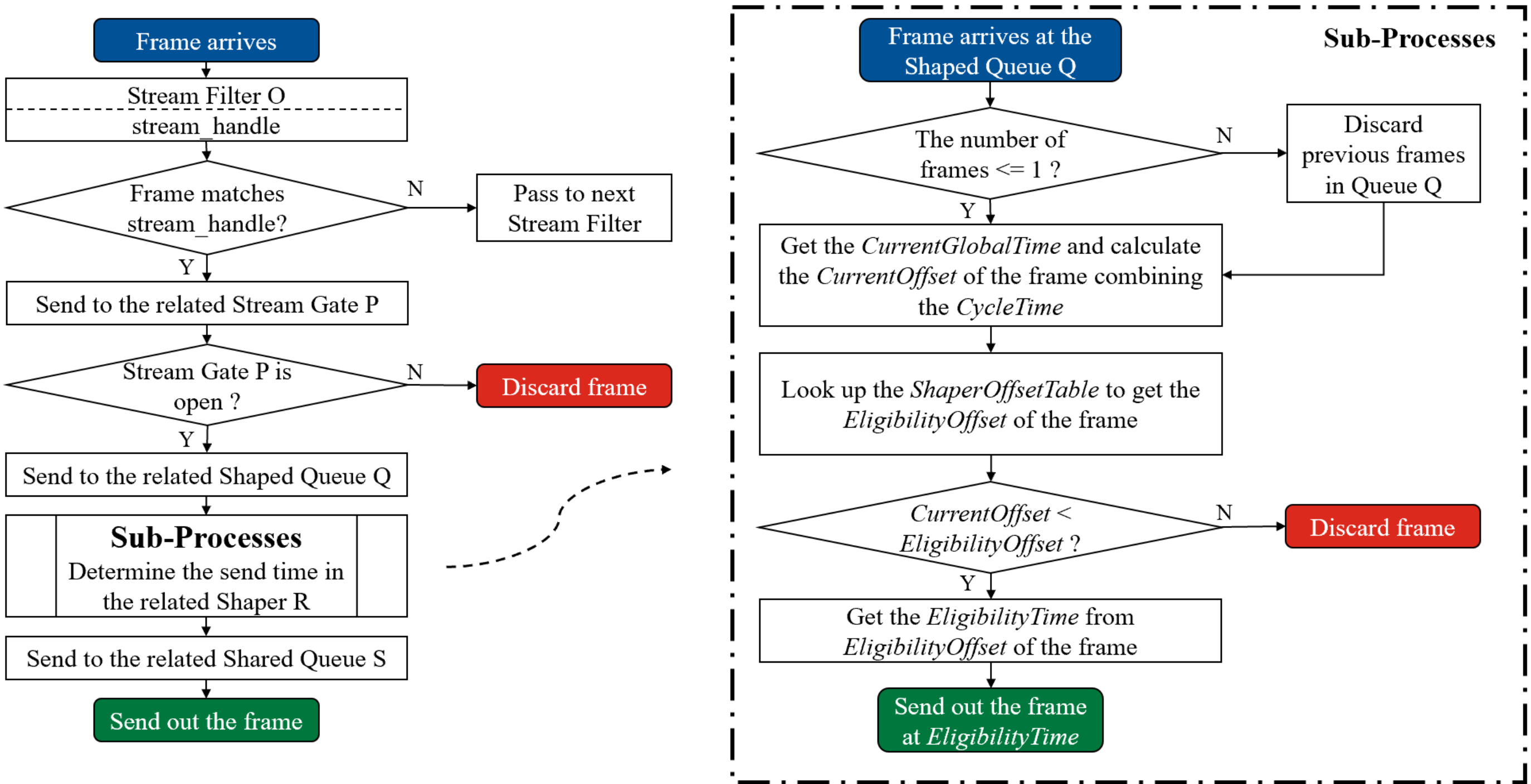

Figure 8. Schematic diagram that illustrates the process that a frame undergoes after entering the TT-UBS mechanism, where the sub-processes show the function of the novel shaper this paper proposed.

Since UBS has a hierarchical queue architecture and the shaping mechanism proposed in this paper specifies the sending time of each frame, rather than controlling the gates of the priority queues as in TAS, the sending order of frames arriving from different ports to the same port can be unaffected by the receiving order even if they are placed in the same shared queue. From the above analysis, it is clear that the frame isolation constraints avoid transmitting jitter caused by the different arrival order or loss of the frames sent from different ports to the same port. Therefore, in the design phase before the mechanism operates, the frame isolation constraints can be removed when the *ShaperOffsetTable* is obtained using the scheduling algorithm, which is analyzed in the next section. Here, we will refer to the algorithm without the frame isolation constraints as the NFIC scheduling algorithm.

### *3.3 Formal Analysis of E2E Latency in the Method*

In this section, we will conduct a formal analysis of the E2E latency experienced by frames during transmission under the proposed mechanism. The key parameters used in the analysis and their respective significance are shown in Table 1.

Table 1 Key parameters in the E2E latency analysis.

| | |
|---|---|
| $L_i$ | The length of frames in the stream i. |
| $r_{tx}$ | The transmission rate of frames in output ports (assuming that the transmission rate of all ports in the network is the same). |
| $D_{i,j}$ | The total delay experienced by the frames of the stream i at the jth hop. |
| $d_{sp,i,j}$ | The delay experienced by the frames of the stream i in the shaped queue at the jth hop. |
| $d_{f,i,j}$ | The forwarding delay experienced by the frames of the stream i at the jth hop. |
| $d_{sr,i,j}$ | The delay experienced by the frames of the stream i in the shared queue at the jth hop. |
| $d_{t,i,j}$ | The transmission delay experienced by the frames of the stream i at the jth hop. |
| $d_{p,i,j}$ | The propagation delay experienced by the frames of the stream i on the link between the jth hop and the (j+1)th hop. |
| $E_{i,j}$ | *EligibilityOffset* of frames in stream i at the jth hop. |

As shown in Figure 9, we refine the delay experienced by stream i at the jth hop into the following components: shaped queue delay, forwarding delay, shared queue delay, transmission delay, and propagation delay. The total delay of stream i at the jth hop can be denoted as Equation (3).

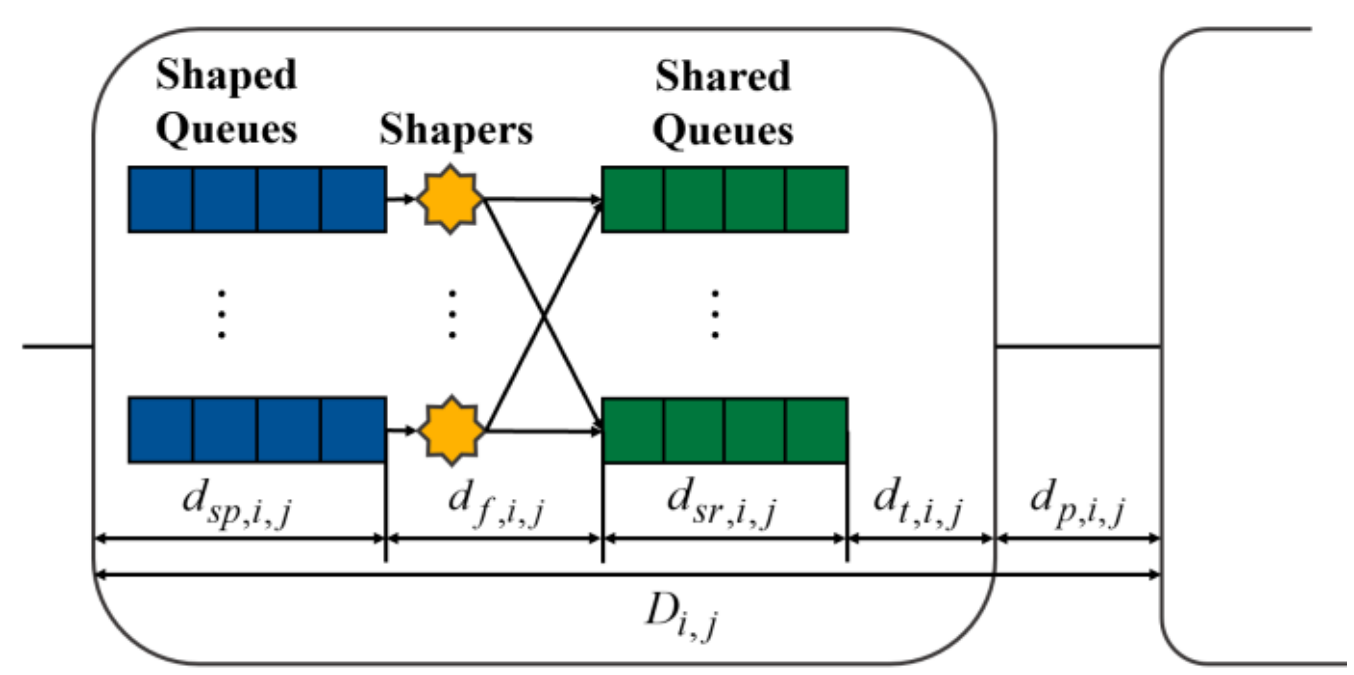

Figure 9. Components of frames E2E latency of in the TT-UBS mechanism in the switch.

$$D_{i,j} = d_{sp,i,j} + d_{f,i,j} + d_{sr,i,j} + d_{t,i,j} + d_{p,i,j} \quad (3)$$

Therefore, for a frame of stream i with a total of m hops, its E2E latency can be represented by Equation (4). Here, $d_{p,i,0}$ represents the transmission delay experienced by the frame from the sending end station.

$$s_i.e2e = d_{p,i,0} + \sum_{j=1}^{m} D_{i,j} \tag{4}$$

Since the forwarding delay and propagation delay of packets are related to the physical characteristics of switches and links and typically in the order of a few clock cycles [31], to simplify the implementation of the simulation in this study, we assume that these parameters ( $d_f$ and $d_p$ ) in the network are fixed to zero. Therefore, the E2E latency of frames can be represented by Equation (5).

$$s_i.e2e = (m+1)d_p + m \cdot d_f + \sum_{j=1}^{m} \left( d_{sp,i,j} + d_{sr,i,j} + d_{t,i,j} \right) \tag{5}$$

Furthermore, since the transmission delay is related to the length of the packet and the transmission rate, the E2E latency of frames in stream i can be represented by Equation (6).

$$s_i.e2e = (m+1)d_p + m \cdot \left( d_f + {L_i}/{r_{tx}} \right) + \sum_{j=1}^{m} \left( d_{sp,i,j} + d_{sr,i,j} \right) \tag{6}$$

From Equation (6), it can be observed that the E2E latency of frames in stream i is mainly influenced by the delay in the shaped queue and shared queue.

For time-critical traffic, the link constraints prevent two frames from being simultaneously transmitted from their respective shaped queues to the same shared queue. This is achieved in the TT-UBS mechanism by leveraging the *EligibilityOffset* of different shaped queues. In addition, best-effort traffic, due to its lower priority, will not be sent to the same shared queue as time-critical traffic. Therefore, in compliance with the constraints, time-critical frames do not experience head-of-line blocking in either the shaped queue or the shared queue during the transmission process. Additionally, due to the constant open gates of shared queues for time-critical traffic, the delay experienced by these frames in the shared queue is zero.

Consequently, the E2E latency of time-critical traffic can be represented by Equation (7). Furthermore, as mentioned above, it is evident that a shaped queue can only hold one frame at one moment. Hence, the delay experienced by a frame in the shaped queue solely depends on its *EligibilityOffset*. In addition, as mentioned above, both the forwarding delay and propagation delay are considered fixed values of zero in this study due to their small numeric values, which are also reflected in the subsequent simulation evaluation process. Consequently, the E2E latency of the frame can also be expressed in terms of the *EligibilityOffset*, as demonstrated in Equation (8). $E_{i,0}$ represents the *EligibilityOffset* of the frame at the port of the sending end station.

$$s_i.e2e = (m+1)d_p + m \cdot \left( d_f + {L_i}/{r_{tx}} \right) + \sum_{j=1}^{m} d_{sp,i,j} \tag{7}$$

$$\begin{aligned} s_i.e2e &= d_{p,i,0} + \left( E_{i,m} - E_{i,0} \right) + d_{f,i,m} + d_{t,i,m} + d_{p,i,m} \\ &= 2d_p + \left( E_{i,m} - E_{i,0} \right) + d_f + {L_i}/{r_{tx}} \\ &= \left( E_{i,m} - E_{i,0} \right) + {L_i}/{r_{tx}} \end{aligned} \tag{8}$$

## *3.4 Performance Analysis of the TT-UBS mechanism*

In this section, we will analyze the performance of the proposed TT communication method under both normal and abnormal traffic conditions, as shown in Figure 10.

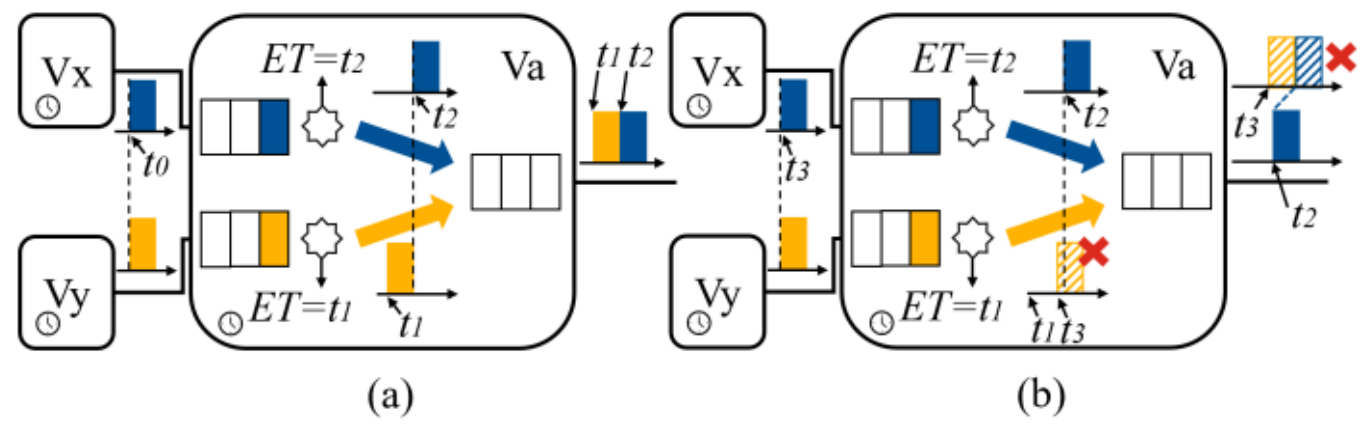

Figure 10. Performance of the TT-UBS mechanism under (a) a normal condition and (b) the condition where a timeout frame exists.

Apply the TT-UBS mechanism to the case shown in Figure 3(a), as shown in Figure 10, where two frames arrive simultaneously and are stored in the same queue (referred to as a shared queue in this method). The analysis of Figure 3(a) reveals that the frame isolation constraints are added because TAS cannot guarantee the definite transmission of frames in this scenario. However, in the TT-UBS mechanism, the dequeuing order is fixed, which allows for the relaxation of the frame isolation constraints.

As shown in Figure 10(a), frames from two end stations arrive at the switch at time t0. After being identified by the Stream Filter component at each switch port, the frames are transmitted to the corresponding shaped queue. They are assigned *EligibilityTime* t1 and t2 by the shaper (assuming t0 is less than t1 and t2). The two frames must then wait in the shaped queue until their *EligibilityTime*, at which point they can be transmitted to the shared queue. Thus, the order in which the frames are transmitted to the output ports is determined by t1 and t2. In this scenario, even if there is an internal forwarding delay in the switch, the order and time at which frames are transmitted are predetermined.

As illustrated in Figure 10(b), similar to Figure 10(a), the shaper assigns *EligibilityTime* t1 and t2 to the two frames. However, in contrast to Figure 10(a), we assume that the two frames arrive at the switch at t3, where t3 falls between t1 and t2. In this scenario, when the shaper detects that the arrival time of a frame exceeds the frame's *EligibilityTime*, it identifies the frame as a timeout frame and discards it to prevent it from affecting the normal transmission of subsequent frames. Figure 10(b) demonstrates that if the shaper does not discard the timeout frame, it will also affect the transmission time of subsequent frames, thereby impacting the traffic transmission in the network.

Similarly, in the event of frame loss, the TT-UBS mechanism in this paper does not affect the transmission of subsequent frames, as they will be sent at their designated times. In the next section, we will evaluate the performance of the mechanism by comparing it with TAS in normal, frame loss, and frame timeout scenarios through simulations. This will allow us to explore the effectiveness of the TT-UBS mechanism in a range of real-world conditions.

## *3.5 The SMT-WA-NFIC Scheduling Algorithm*

The work in [20] proposed the SMT-WA algorithm and solved for the GCL design of the TAS mechanism by designing constraints such as satisfying the physical layer and normal transmission as inputs to the SMT solver. The result solved in this study allows frames to wait for transmission in the queue, as a way of guaranteeing the determinism of frame transmission. The constraints involved in this approach include the individual constraints shown in Equations (9) - (12) and the frame isolation constraints shown in Equation (2).

In this paper, we use Microsoft Z3 Solver, an SMT-based solver, in combination with the constraints above to obtain an appropriate deployment solution. Since the scheduling algorithm proposed in this paper has no frame isolation constraints (NFIC) compared to the SMT-WA scheduling method proposed in [20], we refer to this scheduling algorithm as SMT-WA-NFIC for short.

The above analysis shows that the scheduling solution for the proposed TT communication method based on UBS architecture does not need to satisfy the frame isolation constraints. Therefore, we construct for the other primary constraints except for the frame isolation constraints, which can fulfill the requirement of the FR-WA scheduling solution. The constraints we construct include the frame constraints, the link constraints, the flow transmission constraints, the E2E constraints, and the frame isolation constraints.

The frame constraints can be denoted as Equation (9), which ensures that the time window of a frame must fit within its period. In other words, one frame must be transmitted entirely before the arrival of the frame of the next period, as shown in Figure 11(a). It also ensures that the offset of the frames takes a value greater than or equal to zero.

$$\begin{aligned}&\forall [v_a,v_b]\in E, \forall s_i \in S^{[v_a,v_b]}, \forall f_{i,j}^{[v_a,v_b]} \in F_i^{[v_a,v_b]}:\\&\left(f_{i,j}^{[v_a,v_b]}.\phi \geq 0\right) \wedge \left(f_{i,j}^{[v_a,v_b]}.\phi \leq f_{i,j}^{[v_a,v_b]}.T - f_{i,j}^{[v_a,v_b]}.L\right)\end{aligned} \tag{9}$$

The link constraints which can be denoted as Equation (10) ensures that time windows on the same link do not overlap, as shown in Figure 11(b). This constraint is determined by the physical characteristics of the Ethernet link that two frames cannot be transmitted simultaneously on the same link.

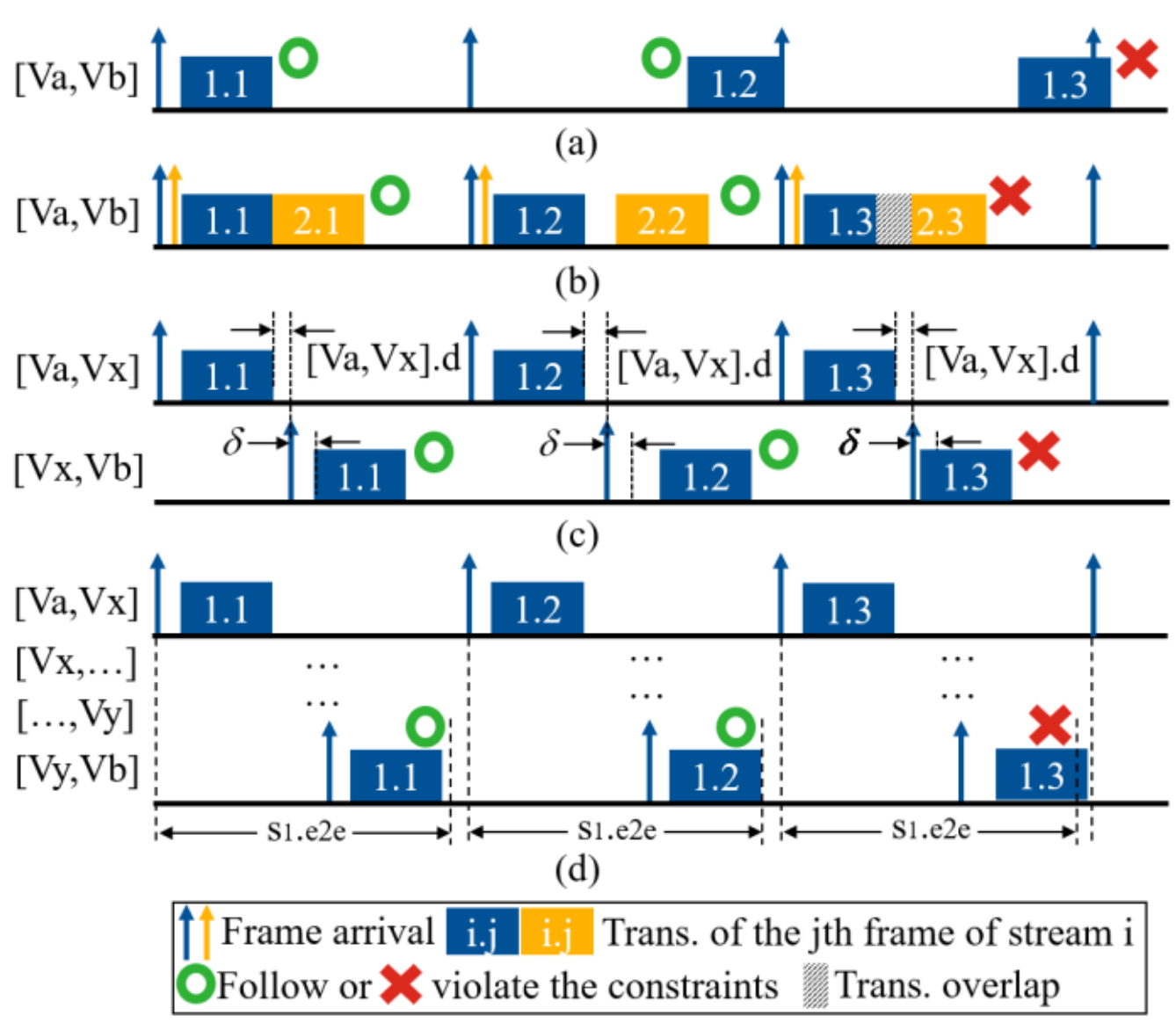

Figure 11. Illustration of (a) the frame constraints, (b) the link constraints, (c) the flow transmission constraints, and (d) the E2E constraints.

$$\begin{aligned}&\forall [v_a,v_b]\in E, \forall s_i, s_j \in S^{[v_a,v_b]}, i \neq j\\&\forall f_{i,k}^{[v_a,v_b]} \in F_i^{[v_a,v_b]}, f_{j,l}^{[v_a,v_b]} \in F_j^{[v_a,v_b]}\\&\forall \alpha \in \left[0, hp^{[v_a,v_b]}/s_i.T-1\right], \forall \beta \in \left[0, hp^{[v_a,v_b]}/s_j.T-1\right]\\&\left(f_{i,k}^{[v_a,v_b]}.\phi + \alpha \cdot f_{i,k}^{[v_a,v_b]}.T \geq\right.\\&\left.f_{j,l}^{[v_a,v_b]}.\phi + \beta \cdot f_{j,l}^{[v_a,v_b]}.T + f_{j,l}^{[v_a,v_b]}.L\right) \vee\\&\left(f_{j,l}^{[v_a,v_b]}.\phi + \beta \cdot f_{j,l}^{[v_a,v_b]}.T \geq\right.\\&\left.f_{i,k}^{[v_a,v_b]}.\phi + \alpha \cdot f_{i,k}^{[v_a,v_b]}.T + f_{i,k}^{[v_a,v_b]}.L\right)\end{aligned} \tag{10}$$

Figure 11(c) shows that the flow transmission constraints guarantee the sequential order in which frames are transmitted over the different links. The time window for a frame on a link should be after its arrival on that link. In Figure 11(c), the variable [va,vb].d represents the propagation delay on the link [va,vb]. It also considers the maximum clock offset $\delta$ between the two devices during synchronization. The flow transmission constraints can be denoted as Equation (11).

$$\begin{aligned}&\forall [v_a,v_x],[v_x,v_b]\in E, \forall s_i \in S^{[v_a,v_x]}, S^{[v_x,v_b]}\\&\forall f_{i,j}^{[v_a,v_x]} \in F_i^{[v_a,v_x]}, \forall f_{i,j}^{[v_x,v_b]} \in F_i^{[v_x,v_b]}\\&f_{i,j}^{[v_x,v_b]}.\phi - \delta \geq\\&f_{i,j}^{[v_a,v_x]}.\phi + f_{i,j}^{[v_a,v_x]}.L + [v_a,v_x].d + [v_a,v_x].p\end{aligned} \tag{11}$$

The E2E constraints which can be denoted as Equation (12) limits the E2E latency of the streams to meet the requirements, as shown in Figure 11(d). The variable s1.e2e in the figure represents the E2E latency requirement of stream 1.

$$\begin{aligned}&\forall [v_a,v_x],[v_y,v_b]\in E, \forall s_i \in S^{[v_a,v_x]}, S^{[v_y,v_b]}\\&\forall f_{i,1}^{[v_a,v_x]}, f_{i,2}^{[v_a,v_x]}, \ldots, f_{i,k}^{[v_a,v_x]} \in F_i^{[v_a,v_x]}\\&\forall f_{i,1}^{[v_y,v_b]}, f_{i,2}^{[v_y,v_b]}, \ldots, f_{i,k}^{[v_y,v_b]} \in F_i^{[v_y,v_b]}\\&f_{i,k}^{[v_y,v_b]}.\phi + f_{i,k}^{[v_y,v_b]}.L - f_{i,1}^{[v_a,v_x]}.\phi \leq s_i.e2e\end{aligned} \tag{12}$$

### *3.6 The AT-NFIC and LS-TB-NIC Algorithm*

Since the TT-UBS mechanism proposed in this paper eliminates the frame enqueue conflicts, the proposed SMT-WA-NFIC algorithm subtracts the frame isolation constraint from the SMT-WA algorithm, as mentioned above. We extend the method to other scheduling algorithms to achieve the solution of the deployment parameters of the TT-UBS mechanism by more scheduling algorithms.

In this section, we modify the scheduling algorithms using Array Theory (AT) [22] and heuristic searching method LS-TB [23] by eliminating the constraints related to the frame enqueue conflicts to obtain the AT-NFIC and LS-TB-NFIC scheduling algorithms. Moreover, we validate the availability of their results and compare their solving time. Note that our abbreviations for the scheduling algorithms are all consistent with those in the review work [13].

The work in [22] expressed the constraints required for TAS deployment via the first-order theory of arrays, and an evaluation using the SMT solver demonstrates the approach's effectiveness. The work shows that it scales well for small to medium-sized networks. The constraints constructed in the AT scheduling algorithm include: well-defined windows constraints, stream instance constraints, ordered windows constraints, frame-to-window assignment constraints, window size constraints, stream constraints, stream isolation constraints, and user constraints, as shown in Figure 12(a). Note that the user constraints here include the maximum transmission latency and jitter constraints to get configuration parameters where both the maximum E2E latency and jitter are met.

The stream isolation constraints are similar to the frame isolation constraints, limiting the time frames being put in the queue. Therefore, we retain the other constraints while deleting the stream isolation constraints to implement the construction of the AT-NFIC scheduling algorithm, as shown by the red dashed circle in Figure 12(a). Like SMT-WA, we use the Z3 solver to solve schemes satisfying these constraints.

The work in [23] points out the limited scalability of using third-party solvers for the solution. It develops an efficient algorithm (LS-TB) for searching based on partial scheduling spaces that do not use any third-party solvers. A brief flow of the LS-TB algorithm is shown in Figure 12(b). After the algorithm assigns offset times and queues to frames, it checks the coordination of the assigned results (checking their compliance with link constraints (LC) and frame isolation constraints). Suppose the frame fails to satisfy these constraints due to conflicts with other frames. In that case, the algorithm jumps to one of the previously allocated frames and tries to allocate a different value, while all frames allocated after that frame are de-allocated.

Since the TT-UBS mechanism avoids frame enqueue conflicts, we skip the checking of frame isolation constraints based on this algorithm (by removing the red dashed box in Figure 12(b)) to obtain the LS-TB-NFIC scheduling algorithm, which enables this search algorithm to perform the backjump operation with lower probability and increase the space for searching.

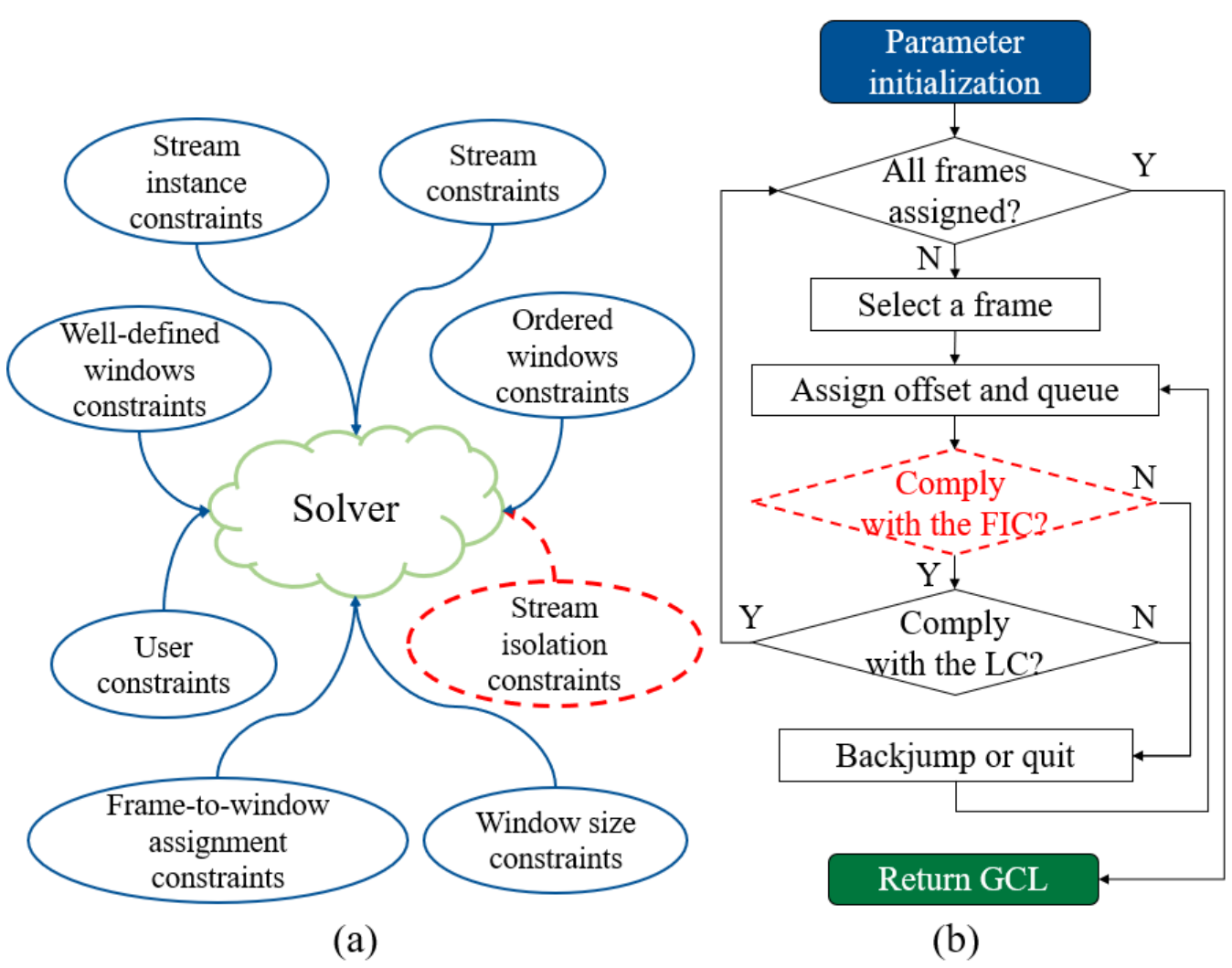

Figure 12. Schematic diagram of (a) AT scheduling algorithm's principle and (b) LS-TB scheduling algorithm's brief flow.

# 4 Simulation Assessment

## *4.1 Simulation Environment and Model Setup*

In this section, the TT-UBS mechanism will be evaluated based on the OMNeT++ simulation environment. Here, to implement TT-UBS, we extended the model's functionality based on the CoRE4INET framework [32], as shown in Figure 13.

As shown in Figure 13(a), we added a submodule called "shapedQueue" to the module that implements the PSFP functionality in the CoRE4INET framework and made partial modifications to other functional submodules such as streamFilter, streamGate, and flow Meter so that frames can be passed through these submodules to the correct shapedQueue module. For convenience of implementation, the shapedQueue submodule integrates the functionality of the TT-UBS mechanism's shaped queues and shapers. On the one hand, this module integrates the functionality of a queue, where arriving frames can be stored in the submodule according to the FIFO principle. On the other hand, the submodule can obtain the arrival time of frames and control the transmission of frames based on their *EligibilityTime*.

In the process of implementing the TT-UBS mechanism, the appropriate modules should be selected. Then, the number of submodules needs to be defined based on the number of streams to be identified. Each Stream Filter component identifies the frame using its priority (also can use its MAC address). This submodule is associated with subsequent submodules by their IDs, such as gateID and meterID so that frames can be passed through them. The newly added shapedQueue submodule only requires two parameters to be configured: ShaperEligibilityOffset and HyperCycleTime. The ShaperEligibilityOffset corresponds to the offset value (unit in seconds) of the frame passing through this submodule in the ShaperOffsetTable. At the same time, HyperCycleTime represents the hyper period of the frames at the output port (unit in microseconds).

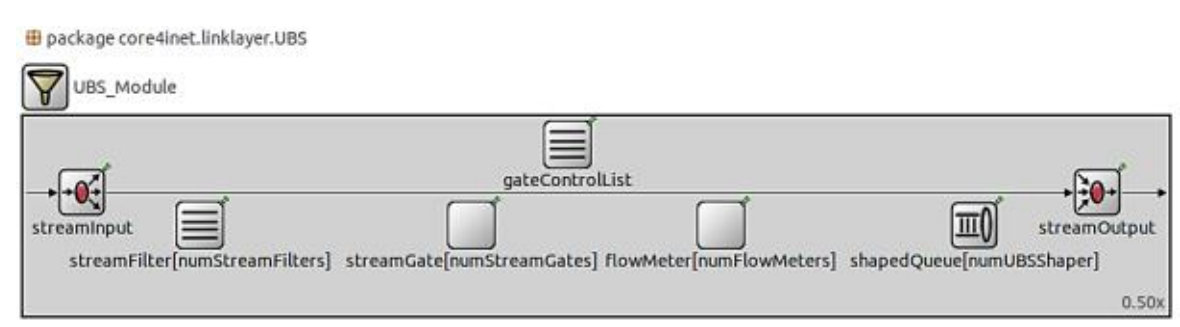

Figure 13. Establishment of the model based on the CoRE4INET framework to implement the TT-UBS mechanism.

Apart from the modifications above, we have developed a dedicated submodule named "AttackMeter" in our simulation to facilitate the creation of scenarios involving frame loss and timeouts. This module serves as a replacement for the flowMeter submodule depicted in Figure 13. It enables us to discard and delay designated frames selectively. After configuring the submodule and specifying the attackType (1 for frame loss, 2 for frame delay), it is essential to determine the submodule's StartupTime (unit in seconds) and frameCount (representing the quantity of discarded frames). The submodule will discard the designated number of frames after the specified start time. Moreover, the configuration that implements the functionality of frame delay introduces an additional parameter called delayTime (unit in seconds). This parameter represents the duration of the frame delay.

### *4.2 Performance under Different Scenarios*

#### 4.2.1 Network and Traffic Model

To evaluate whether the proposed UBS-based approach can meet the requirements, we utilize the star topology Advanced Driver Assistance Systems (ADAS) fusion zone system proposed in [33] for automotive applications. The ADAS fusion zone system with the star topology consists of two switches and five nodes, where the Central Host with a switch makes up the Vehicle Computation Unit (VCU) and the Zonal Host with a switch makes up the Vehicle Interface Unit (VIU). The difference is that we ramp up the link rate to 1Gbps and define the traffic according to Table 2, as shown in Figure 14.

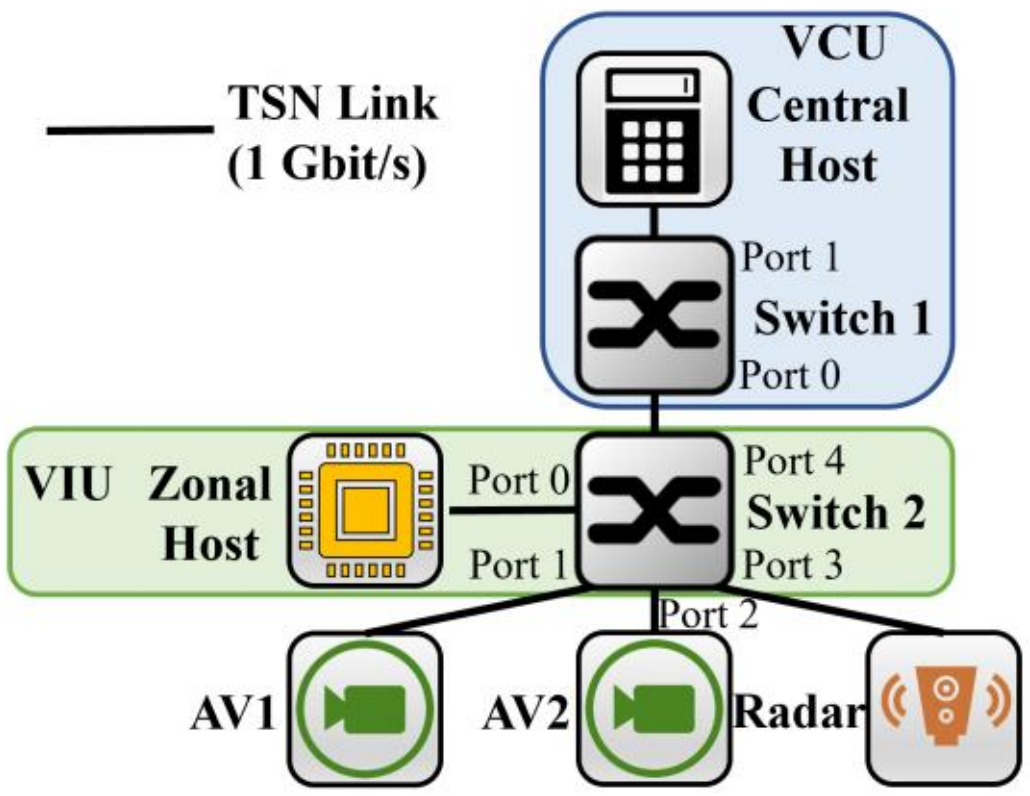

Figure 14. ADAS fusion zone system with the star topology.

Table 2. Parameters of the traffic transmitted in the ADAS fusion zone system with the star topology.

| Number | 1 | 2 | 3 | 4 |
|---|---|---|---|---|
| Stream Info. | Cam. data 1 | Cam. data 2 | Radar data | Ctrl. data |
| Source | AV1 | AV2 | Radar | Zonal Host |
| Destination | Ctr. Host | Ctr. Host | Ctr. Host | Ctr. Host |
| Payload (B) | 1000-1200 | 1000-1200 | 300-400 | 150-200 |
| Period (us) | 100 | 100 | 200 | 200 |
| Ddl. (us) | 100 | 100 | 200 | 200 |
| Jt. Req. (us) | 10 | 10 | 20 | 20 |

In the scenario depicted in Figure 14, camera data is transmitted from AV1 and AV2 to the Central Host, radar data is transmitted from Radar to the Central Host, and control data is transmitted from the Zonal Host to the Central Host. As a frame not received beyond its sending period loses its timeliness, the deadline for each stream is assumed to be its sending period, with the jitter requirement set at 10% of the sending period.

#### 4.2.2 Deployment Schemes

To prove that the TT-UBS mechanism can work properly under the deployment scenarios obtained by the SMT-WA-NFIC scheduling algorithm, we obtained the deployment solution of the TT-UBS mechanism using the SMT-WA-NFIC scheduling algorithm. In addition, we deployed the results obtained using this scheduling algorithm under the TAS mechanism as a comparison to show that this scheduling algorithm is not applicable to the TAS mechanism. In other words, the lack of frame isolation constraints can seriously impact the effectiveness of TAS.

These solutions (ShaperOffsetTable and GCL of port 4 of switch 2) are shown in Tables 4 and 5, respectively. In addition, the first frame of each stream in a hyper-period has a sending time of zero at the port of its sending device, which means that $E_{i,0}$ of the first frame of the stream i in a hyper-period equals zero. If the hyper-period is different from the period of a stream, then the values of $E_{i,0}$ for the subsequent frames in that stream increase by multiples of the period of the stream. For instance, the second frames of Camera Data 1 and Camera Data 2 during a hyper-period have an $E_{i,0}$ of 100us. It should be noted that the Z3 solver evaluates whether the current scenario satisfies the requirements of the Boolean Satisfiability Problem (SAT) before executing the solving process. Consequently, it will not yield any results if the specified requirements are not met. Therefore, all the solutions presented in the validation section of this research paper are obtained under the case that the input satisfies the SAT requirements.

From Tables 4 and 5, it can be observed that the frames sent at switch 2 have the same offset. This is because their deployment schemes are derived from the same algorithm result. It should be noted that, due to the removal of the frame isolation constraints, all frames are allocated to the same queue. As a result, in Table 4, the GCL demonstrates that only the queue with a priority of 4 undergoes repeated opening and closing to regulate the transmission of these frames. In contrast, other gates can be configured to open at different intervals. The solution was obtained in a mere 0.0049998 s, which is obtained by running the algorithm on a computer equipped with a 12th Gen Intel Core i5-12400 64-bit 2.50GHz CPU and 16GB of RAM.

Table 3. The ShaperOffsetTable obtained by the SMT-WA-NFIC scheduling algorithm (unit in us).

| Switch | Port | Stream | *EligibilityOffset* | *CycleTime* |
|---|---|---|---|---|
| Switch 2 | Port 0 | Control data | 3 | 200 |
| | Port 1 | Camera data 1 | 21 | 200 |
| | | | 121 | |
| | Port 2 | Camera data 2 | 11 | 200 |
| | | | 111 | |
| | Port 3 | Radar data | 5 | 200 |
| Switch 1 | Port 0 | Control data | 6 | 200 |
| | | Camera data 1 | 32 | 200 |
| | | | 132 | |
| | | Camera data 2 | 22 | 200 |
| | | | 122 | |
| | | Radar data | 10 | 200 |

Table 4. The GCL of port 4 on switch 2, with a cycle time of 200 us, is obtained using the SMT-WA-NFIC scheduling algorithm ('C' represents closing and 'o' represents open).

| Interval (us) | Q0 | Q1 | Q2 | Q3 | Q4 | Q5 | Q6 | Q7 |
|---|---|---|---|---|---|---|---|---|
| 0-3 | o | o | o | o | C | o | o | o |
| 3-4 | C | C | C | C | o | C | C | C |
| 4-5 | o | o | o | o | C | o | o | o |
| 5-7 | C | C | C | C | o | C | C | C |
| 7-11 | o | o | o | o | C | o | o | o |
| 11-19 | C | C | C | C | o | C | C | C |
| 19-21 | o | o | o | o | C | o | o | o |
| 21-29 | C | C | C | C | o | C | C | C |
| 29-111 | o | o | o | o | C | o | o | o |
| 111-119 | C | C | C | C | o | C | C | C |
| 119-121 | o | o | o | o | C | o | o | o |
| 121-129 | C | C | C | C | o | C | C | C |
| 129-200 | o | o | o | o | C | o | o | o |

### 4.2.3 The Normal Traffic Case

After obtaining the deployment solutions for the TT-UBS mechanism and TAS using the SMT-WA-NFIC scheduling algorithm, we evaluated their performance in normal scenarios. To this end, we employed OMNeT++ simulation to obtain the E2E latency and jitter for each stream in the network. The simulation results are presented in Figure 15, which compares TT-UBS and TAS regarding their E2E latency and jitter. It is important to note that the duration of all simulation runs in this paper has been set to 10 seconds, significantly longer than the maximum period for frame transmission.

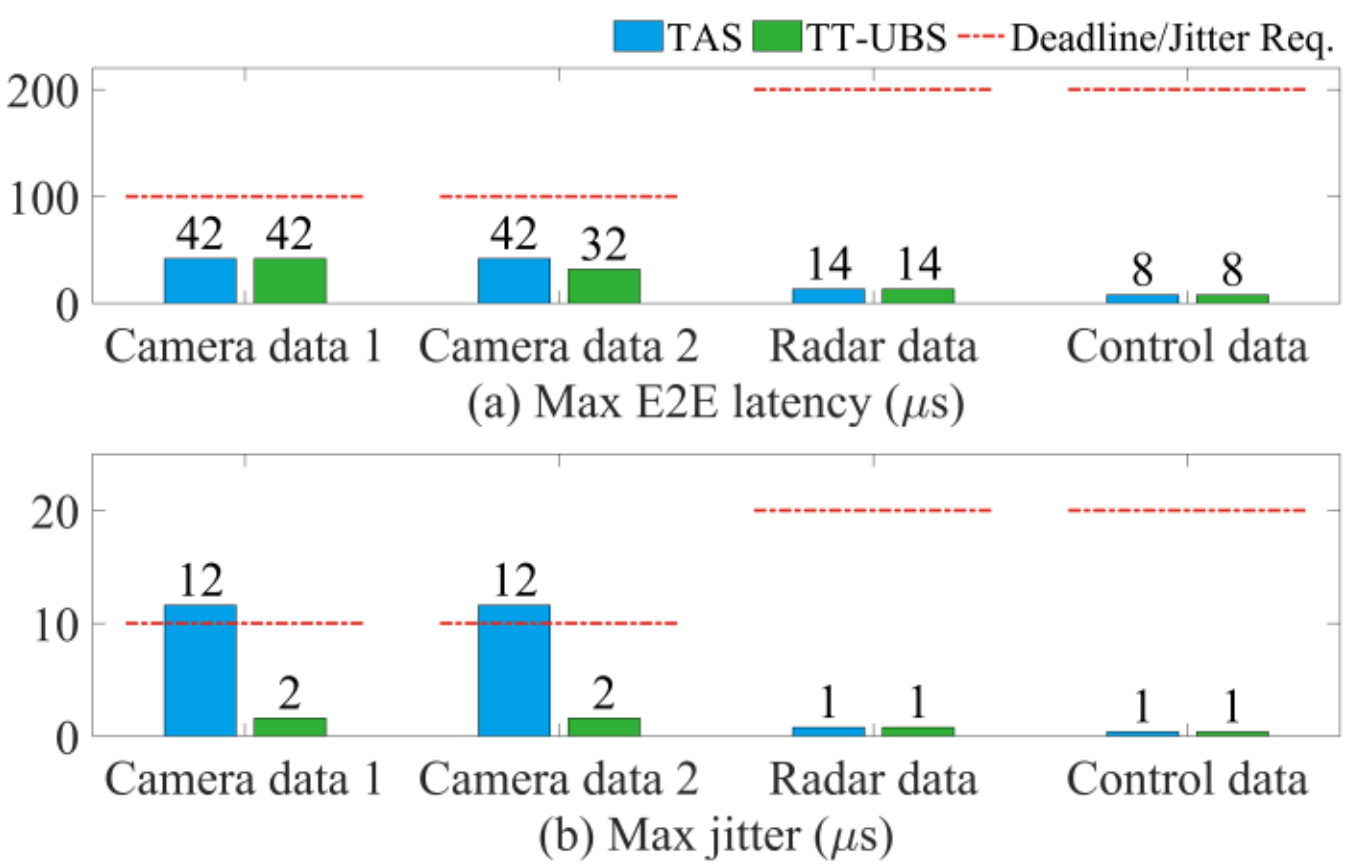

Figure 15. (a) Max E2E latency and (b) Max jitter of streams using TT-UBS and TAS, whose solutions are obtained by the SMT-WA-NFIC scheduling algorithm under the normal case.

Based on the analysis of Figure 15(a), it is evident that in the normal case when the TAS is applied, both Camera data streams exhibit a maximum E2E latency of 42us. The Radar data stream experiences a maximum E2E latency of 14us, while the control data stream has a maximum E2E latency of 8us. When implementing the TT-UBS mechanism, it is observed that all streams, except for one Camera data stream, maintain the same maximum E2E latency as observed when TAS is applied. This shows that when employing the SMT-WA-NFIC scheduling algorithm, TT-UBS performs comparably to TAS by successfully fulfilling the E2E latency requirements for various traffic types within the network under the normal case. However, Figure 15(b) indicates that when TAS is applied, neither of the two Camera data streams meets their respective jitter requirements. In contrast, when TT-UBS is implemented, all traffic streams can meet the jitter requirements. This is primarily attributed to removing the frame isolation constraints, which results in the non-deterministic order of the two Camera data streams entering the queue. Consequently, this introduces significant transmission jitter of these streams.

The results in Figure 15 also serve as evidence for our previous formal description of the E2E latency for streams. Taking the first stream Camera Data 1 as an example, it sends two frames within one hyper-period with an interval of 100us. For the first frame, its *EligibilityOffset* at the first link is zero (as mentioned earlier, $E_{1,0}$ is 0 in this case). Since the frame undergoes a two-hop transmission to reach its receiving device, $E_{1,2}$ corresponds to the *EligibilityOffset* at Switch 2 for that frame. Therefore, according to Table 3, $E_{1,2}$ is equal to 32us. In addition, we can infer from Table 2 that the maximum payload size of the frame is 1200 bytes, while the minimum size is 1000 bytes. Therefore, when the payload is at its maximum size, the length of the frame is the payload length plus an additional 22 bytes for frame headers, checksum, and other content, resulting in a total length of 1222 bytes. Similarly, the shortest length of the frame is 1022 bytes. The maximum E2E latency can be expressed using Equation (13) and the minimum E2E latency can be expressed as Equation (14). Jitter can be obtained by calculating the difference between the maximum E2E latency and the minimum E2E latency, as shown in Equation (15).

$$s_1.e2e_{\max} = \left(E_{1,2} - E_{1,0}\right) + \max\left(L_1\right) / r_{tx} = \left(\left(32 - 0\right) + 1e6 \times 8 * 1222 / 1e9\right) us \approx 42us \tag{13}$$

$$s_1.e2e_{\min} = \left(E_{1,2} - E_{1,0}\right) + \min\left(L_1\right) / r_{tx} = \left(\left(32 - 0\right) + 1e6 \times 8 * 1022 / 1e9\right) us \approx 40us \tag{14}$$

$$s_1.jitter = s_1.e2e_{\max} - s_1.e2e_{\min} = 42us - 40us = 2us \tag{15}$$

Similarly, for the second frame of this stream in a hyper-period, $E_{1,0}$ is 100us (as mentioned earlier). From Table 3, it can be seen that $E_{1,2}$ is equal to 132us. From the above formulas, it can be concluded that the maximum E2E latency and jitter of the second frame of the stream during transmission within a hyper-period are the same as those of the first frame.

The calculation of E2E latency for other streams follows a similar approach. By utilizing formulas (13), (14), and (15), we can demonstrate our previous formal description of the E2E latency.

### 4.2.4 The Abnormal Traffic Cases

To construct the cases of frame loss and timeout mentioned above, we configured the parameters of the AttackMeter submodule, as shown in Table 5.

Table 5. Parameter configuration of the AttackMeter submodule.

| Scenario | Switch | Port | Parameter | Value |
|---|---|---|---|---|
| Frame loss | Switch 2 | Port 1 | attackType | 1 |
| | | | startupTime | 21 us |
| | | | frameCount | 1 |
| Frame timeout | Switch 1 | Port 0 | attackType | 2 |
| | | | startupTime | 21 us |
| | | | delayTime | 10 us |
| | | | frameCount | 1 |

Based on the findings presented in Figure 16, it is clear that when faced with frame loss, both TAS and TT-UBS exhibit performance similar to that of the normal case. The loss of a single frame does not impact the subsequent frames. This behavior can be attributed to allocating specific time windows for each frame using TAS, as described in this paper. This allocation ensures that there are no consecutive frames transmitted within a single time window, thereby mitigating the impact of frame loss.

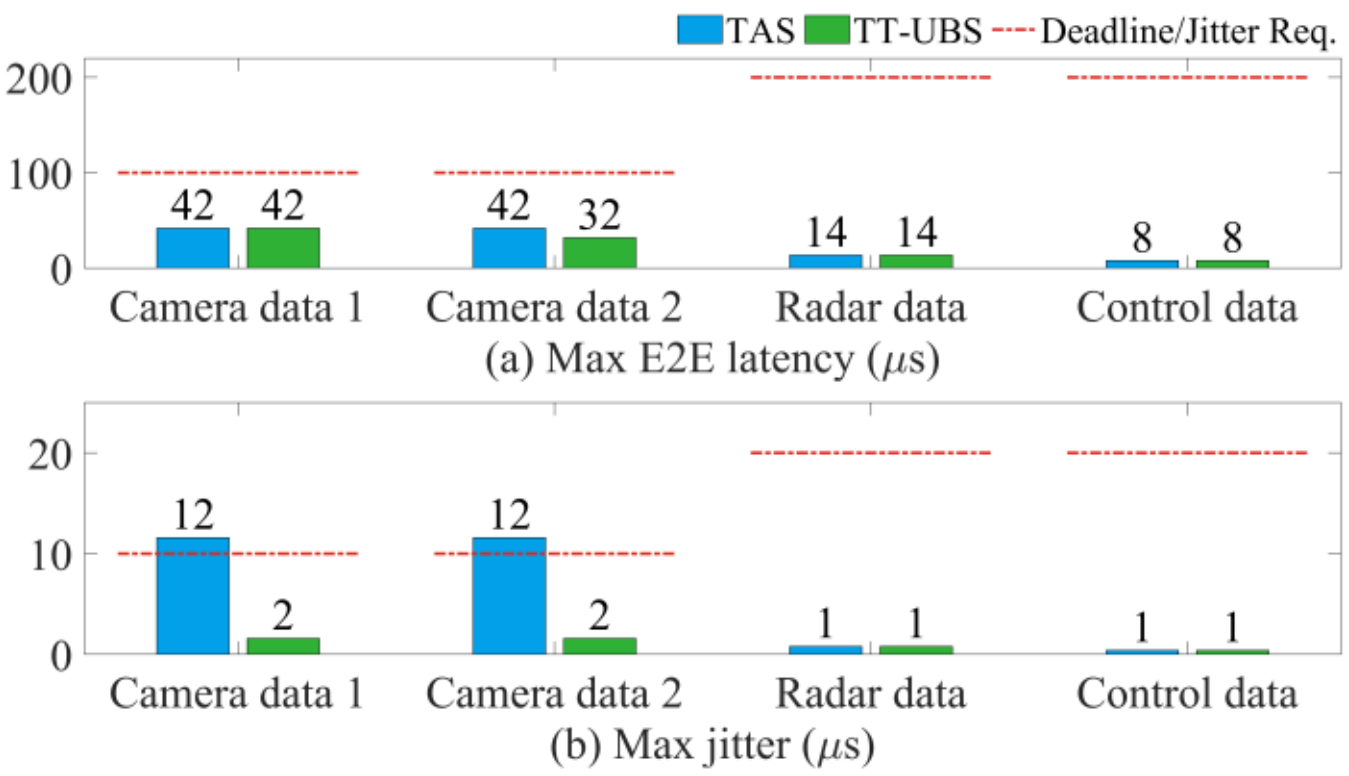

Figure 16. (a) Max E2E latency and (b) Max jitter of streams using TT-UBS and TAS, whose solutions are obtained by the SMT-WA-NFIC scheduling algorithm under the frame loss case.

Figure 17 presents the maximum E2E latency and jitter of the streams under a frame delay scenario of 10us for Camera data 2. Figure 16 demonstrates that in the case of frame timeout, using TAS leads to an increase in the maximum E2E latency and jitter for most traffic streams, with some even exceeding their transmission periods (Camera data 2). This outcome results in the loss of effectiveness for these frames. Furthermore, these findings further validate the analysis presented in Figure 6. However, for the TT-UBS mechanism, the impact of the frame timeout on the E2E latency and jitter of streams can be considered negligible. This is because the TT-UBS mechanism discards timed-out frames to prevent them from interfering with the transmission of subsequent frames. This further demonstrates that the TT-UBS mechanism exhibits good performance in scenarios with abnormal traffic.

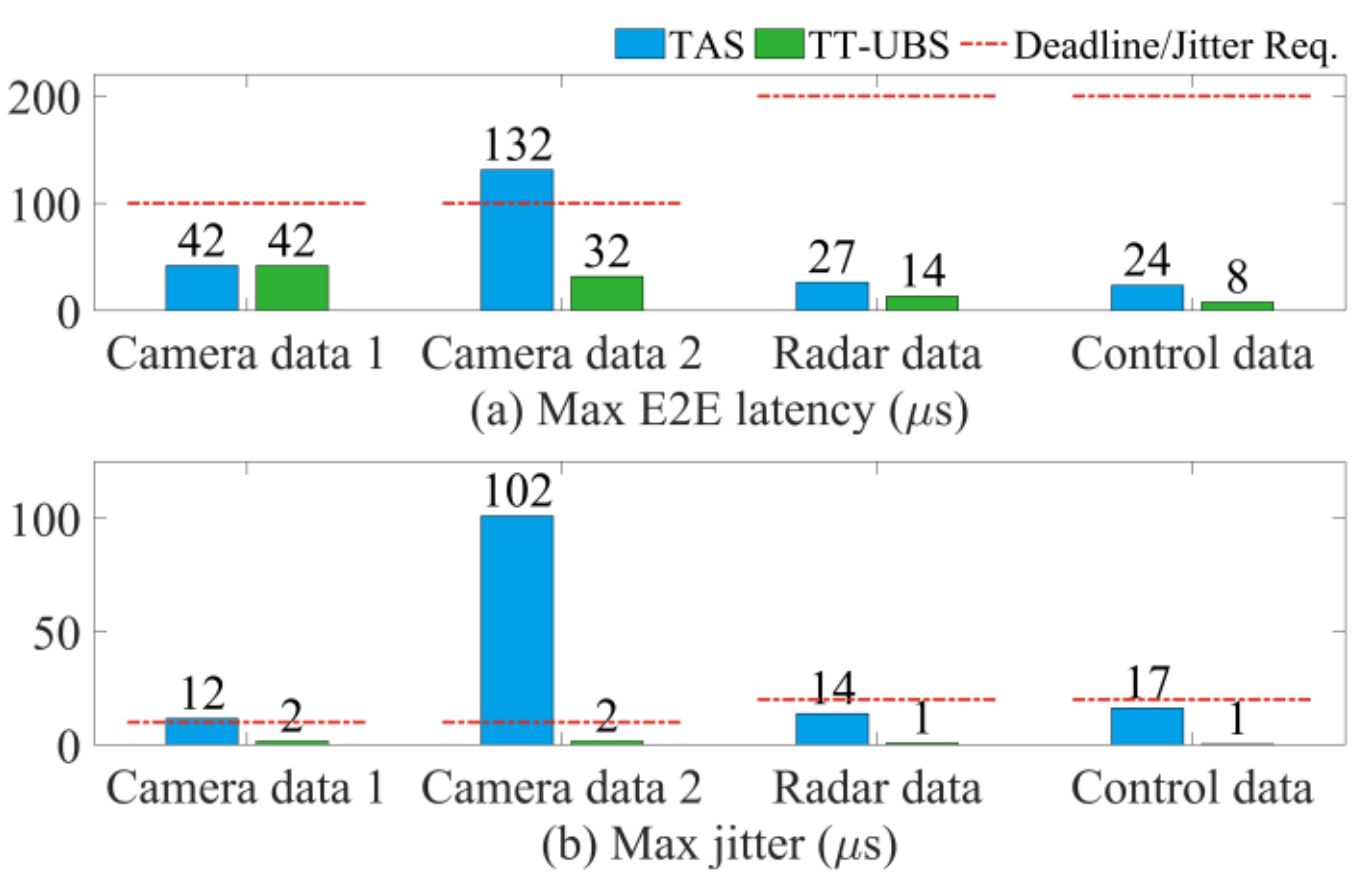

Figure 17. (a) Max E2E latency and (b) Max jitter of streams using TT-UBS and TAS, whose solutions are obtained by the SMT-WA-NFIC scheduling algorithm under the timeout case.

The above simulation evaluation shows that the results derived from the proposed SMT-WA-NFIC scheduling algorithm can be better deployed on the TT-UBS mechanism. However, the algorithm cannot be applied to the TAS mechanism because the frames must avoid the enqueue conflict, which the TT-UBS mechanism can avoid.

## *4.3 Compare with SMT-WA scheduling algorithm*

To evaluate the proposed SMT-WA-NFIC scheduling algorithm, we compare the results and solution speeds obtained by this algorithm with the SMT-WA scheduling algorithm proposed in [20].

### 4.3.1 Comparison of Solving Results

To facilitate a more comprehensive comparison, this section examines the performance of the TT-UBS mechanism in scenarios where the delay exceeds the transmission period. In addition, this section explores the results obtained by the proposed SMT-WA-NFIC scheduling algorithm and the results obtained using the SMT-WA scheduling algorithm as implemented in TT-UBS and under TAS, respectively.

We set the delayTime in Table 5 to 221 us, which implies that the frames transmitted by Camera 2 are delayed by more than one cycle time. The result is shown in Figure 18.

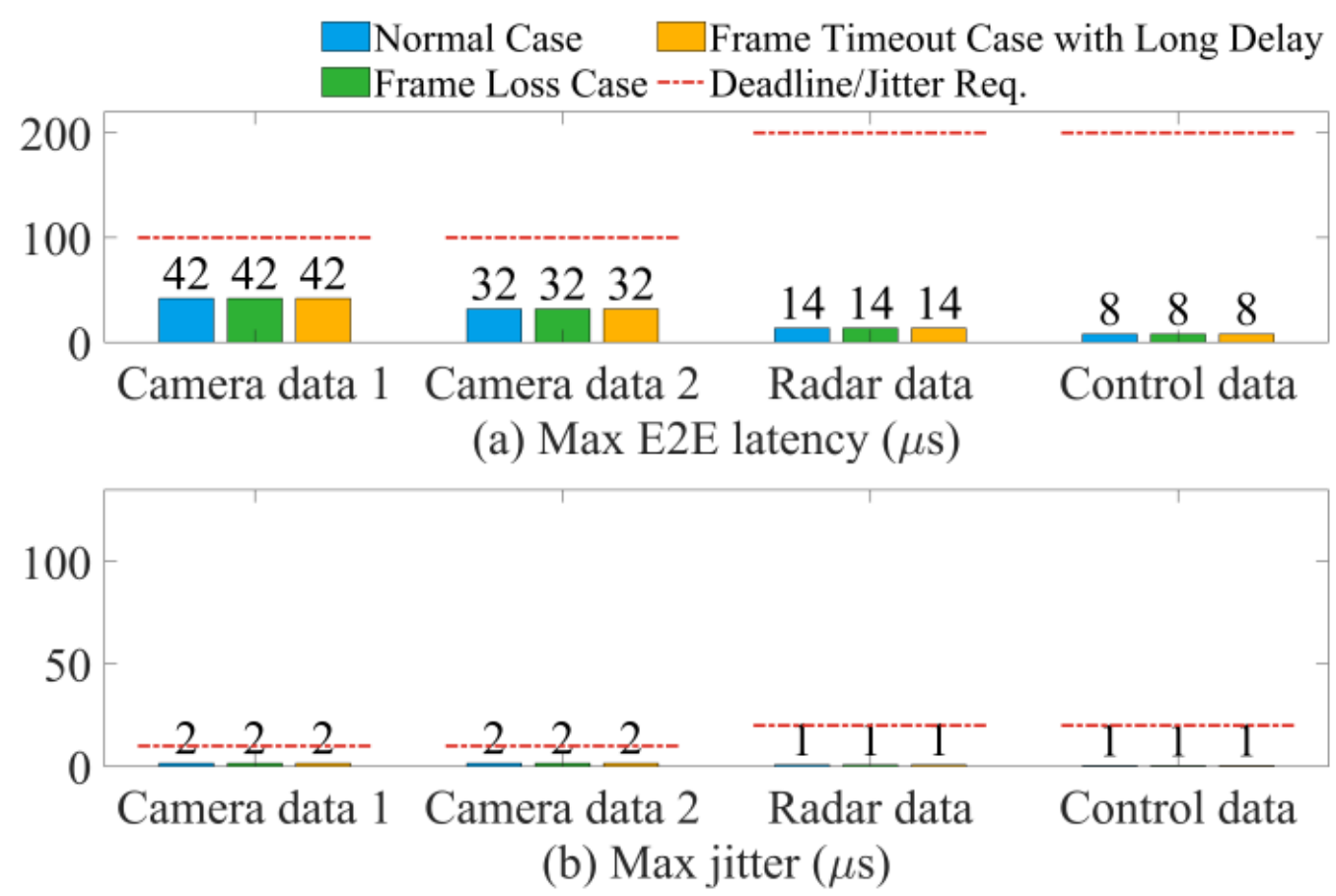

Figure 18. (a) Max E2E latency and (b) Max jitter of streams using TT-UBS whose solutions are obtained by the SMT-WA-NFIC scheduling algorithm under the normal case, frame loss case, and frame timeout case with long delay.

Figure 18 provides evidence that the TT-UBS mechanism maintains the required E2E latency of frames under abnormal traffic scenarios, even when the delay exceeds the cycle time. In such cases, multiple timed-out frames may accumulate in a shaped queue. To ensure the unhindered transmission of subsequent frames, the TT-UBS mechanism discards these timed-out frames in the shaped queue. This indicates that even in the presence of long delays, the TT-UBS mechanism maintains the transmission quality of traffic in the network.

Moreover, we explore the deployment scheme of TAS obtained using the SMT-WA scheduling algorithm proposed in [20]. Part of the solution is shown in Table 6. It is worth noting that the introduction of the frame isolation constraints in SMT-WA results in an increase in the execution time on the same device, from 0.0049998 s to 0.0185155 s.

Table 6. The GCL of port 4 of switch 2 obtained by the SMT-WA scheduling algorithm.

| Interval (us) | Q0 | Q1 | Q2 | Q3 | Q4 | Q5 | Q6 | Q7 |
|---|---|---|---|---|---|---|---|---|
| 0-3 | o | o | o | o | C | C | C | C |
| 3-4 | C | C | C | C | C | C | C | o |
| 4-5 | o | o | o | o | C | C | C | C |
| 5-7 | C | C | C | C | C | C | o | C |
| 7-11 | o | o | o | o | C | C | C | C |
| 11-19 | C | C | C | C | C | o | C | C |
| 19-21 | o | o | o | o | C | C | C | C |
| 21-29 | C | C | C | C | o | C | C | C |
| 29-111 | o | o | o | o | C | C | C | C |
| 111-119 | C | C | C | C | C | o | C | C |
| 119-121 | o | o | o | o | C | C | C | C |
| 121-129 | C | C | C | C | o | C | C | C |
| 129-200 | o | o | o | o | C | C | C | C |

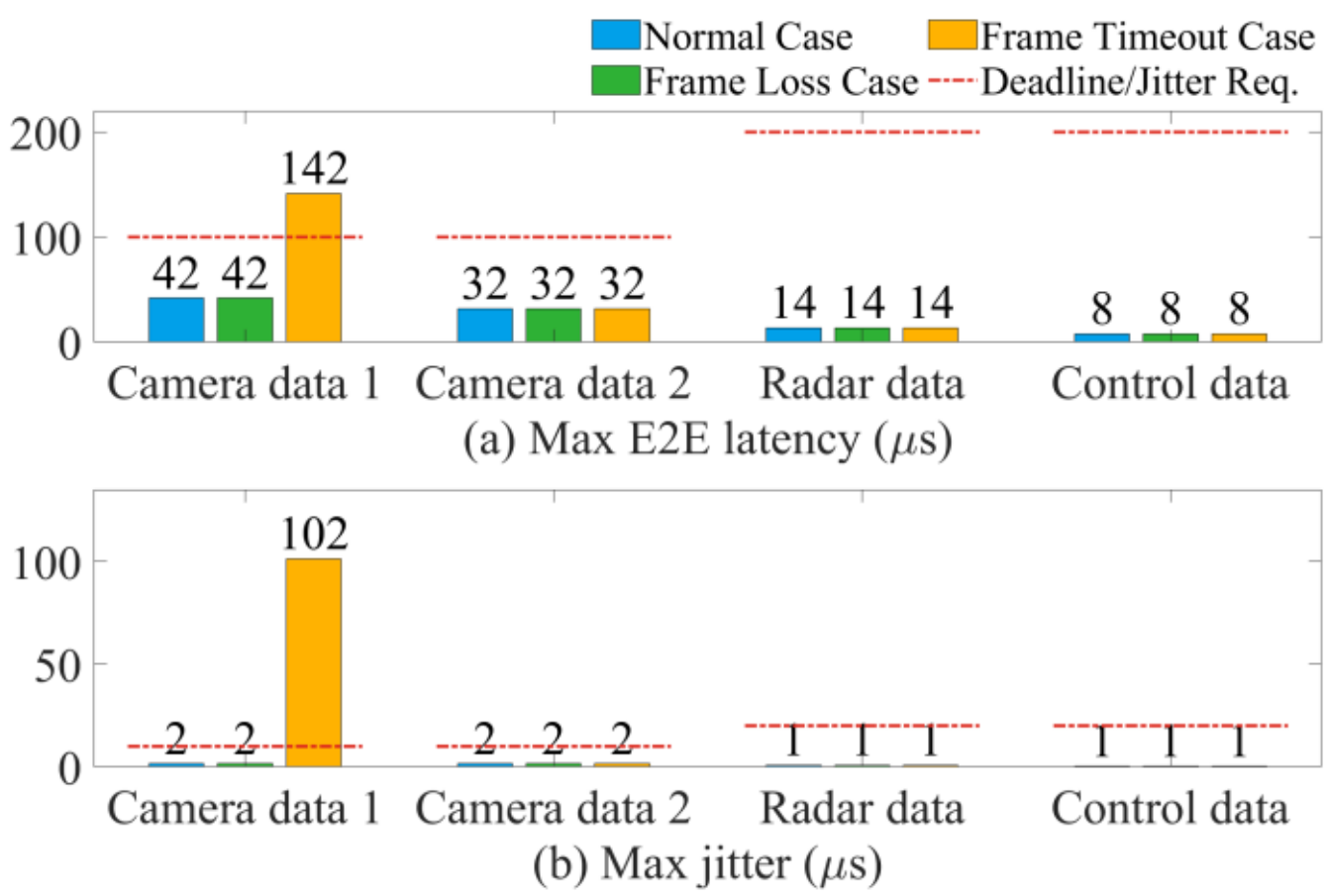

Figure 19. (a) Max E2E latency and (b) Max jitter of streams using TAS, whose solution is obtained by the SMT-WA scheduling algorithm under the normal case, frame loss case, and frame timeout case.

Based on Figure 19, it is evident that when using the SMT-WA scheduling algorithm to obtain the solution, TAS can meet the deadline and jitter requirements in normal and frame loss cases. However, it still fails to meet the requirements regarding frame timeouts. This demonstrates that the TT-UBS mechanism outperforms TAS in cases with abnormal traffic, highlighting its superiority in handling such situations.

### 4.3.2 Comparison of Solving Speed

This section compares the SMT-WA-NFIC and SMT-WA scheduling algorithms, mainly regarding solution speed. We use the topology in Figure 5 with the number of devices ranging from 4 to 36. Similarly, we increase the number of streams from 5 to 95 in steps of 5 to compare the variation in solving time of the SMT-WA-NFIC and SMT-WA scheduling algorithms. Similarly, all experiments were run on a computer equipped with a 12th Gen Intel core i5-12400 64-bit 2.50GHz CPU with 16GB RAM. We repeated the solving process 50 times and averaged the results for each iteration. To enhance the solving efficiency, we considered solving instances with a duration exceeding 5 minutes as ineffective. Therefore, we set a time limit of 5 minutes for each solve instance. If the solving time exceeded this limit, we considered it unable to find a solution and forcefully terminated the solving process. The variation of solving time under these two algorithms is shown in Figure 20.

As seen in Figure 20, the solving time under the two algorithms increases overall with the stream number increase. Figure 20(a) demonstrates that the solving time of the SMT-WA algorithm does not increase with the increase in the number of network devices in a scenario where the number of streams is the same. This is because there are times when an increase in the number of devices may lead to a decrease in the number of traffic conflicts (less likelihood of sharing the same link for transmission), which leads to a reduction in the number of constraints, and hence a decrease in the solving time.

Figures 20(b)-(f) show the variation of the solving time with the increase in the number of streams for the SMT-WA and SMT-WA-NFIC algorithms for different numbers of devices, respectively, which shows that the solving time of SMT-WA is always greater than that of the SMT_NFIC algorithm, no matter what the number of devices and streams are. The difference in the solving time between them is insignificant when the number of streams is small (e.g., in the case where the number of streams is 5 and the number of devices is 36 in Figure 20(f)), and the difference increases as the number of streams increases.

Figure 20 also shows that the solving time under SMT-WA shows a substantial increase (from within 1 second to tens of seconds) when the number of streams is in the 45 to 65 range. The reasons behind this phenomenon can be attributed to the increase in the number of constraints on one hand and the characteristics of the solver on the other hand (reaching a certain threshold of constraints may lead to a significant increase in solving time). However, the solving time under the SMT-WA-NFIC algorithm is more stable and is always smaller than that under the SMT-WA algorithm. The SMT-WA-NFIC algorithm reaches the maximum solving time of about 0.7 seconds, with the number of devices being 36 and the number of streams being 90. At the same time, the SMT-WA algorithm achieves the maximum solving time of about 41 seconds for the same case. In addition, the solving time of the SMT-WA-NFIC algorithm does not increase stepwise as the network expands but increases more smoothly.

From the above analysis, it can be seen that the use of the SMT-WA-NFIC scheduling algorithm can reduce the solving time, which means that when the TT-UBS mechanism is used in the network, a feasible solution can be solved by using the SMT-WA-NFIC scheduling algorithm, which can effectively reduce the solving time.

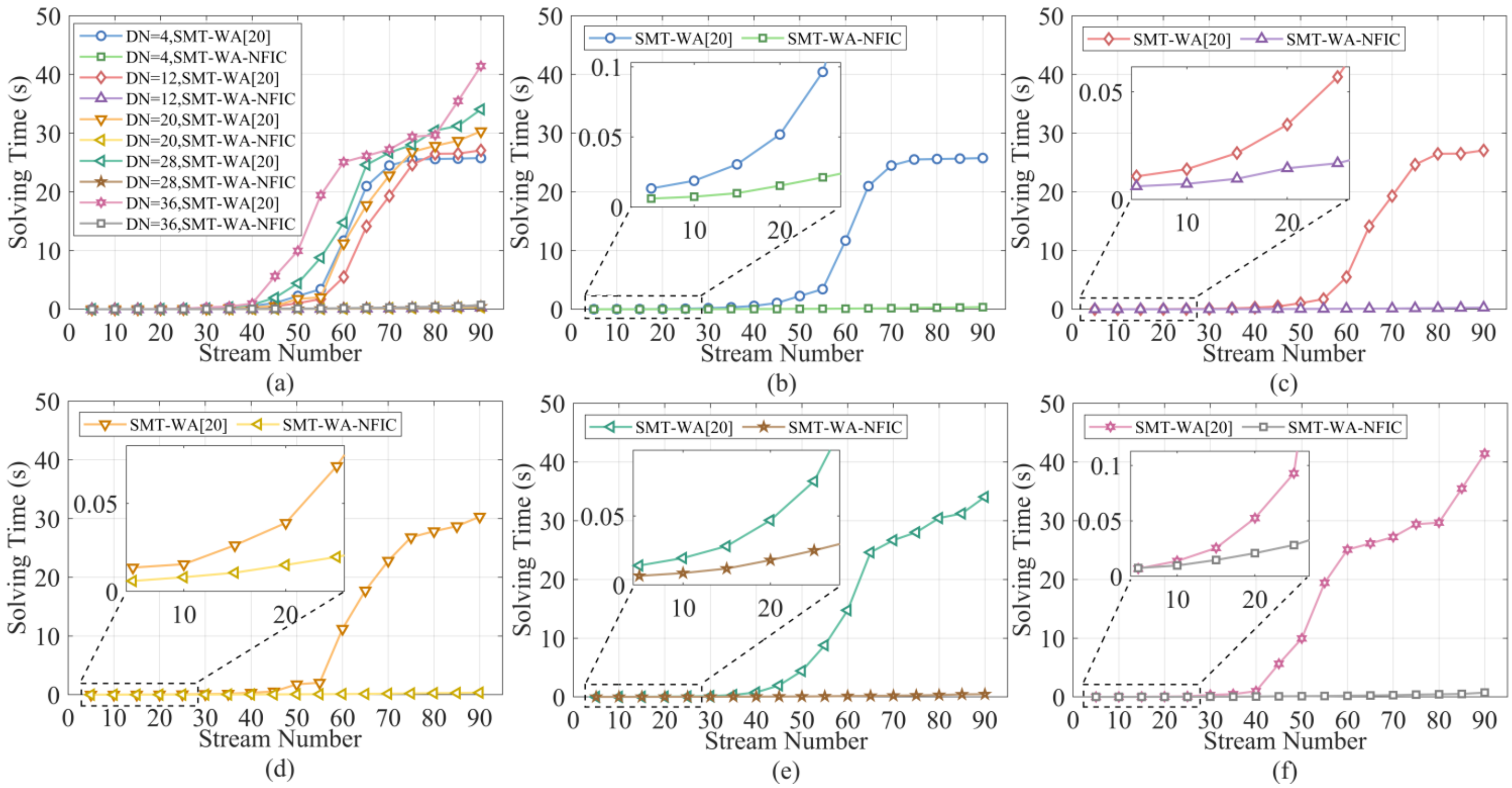

Figure 20. The variation of solving time with the increase of stream number under the SMT-WA-NFIC and SMT-WA scheduling algorithms. (a) All cases, (b) Device Number is 4, (c) Device Number is 12, (d) Device Number is 20, (e) Device Number is 28, and (f) Device Number is 36.

## *4.4 Compare with AT and LS-TB*

### 4.4.1 Availability of Solving Results

To verify that the results solved by our modified scheduling algorithms can satisfy the transmission requirements of the frames, we deploy the results of their solution in software simulation in this section, and the adopted network is used in the application scenario related to automotive ADAS as previously described.

We take the topology shown in Figure 13 and the traffic attributes shown in Table 2 as inputs and obtain the frame's offset time on each link through the AT-NFIC algorithm, which is then transformed into parameters related to the TT-UBS mechanism (*EligibilityOffset*) as shown in Table 7. Note that in the AT(-NFIC) scheduling algorithm, the maximum number of GCL entries is also an input parameter to specify the maximum number of windows in the result. In this paper, the value is taken as 4.

Table 7. The ShaperOffsetTable obtained by the AT-NFIC scheduling algorithm (unit in us).

| Switch | Port | Stream | *EligibilityOffset* | *CycleTime* |
|---|---|---|---|---|
| Switch 2 | Port 0 | Control data | 74 | 200 |
| | Port 1 | Camera data 1 | 64 | 200 |
| | | | 137 | |
| | Port 2 | Camera data 2 | 31 | 200 |
| | | | 151 | |
| | Port 3 | Radar data | 147 | 200 |
| Switch 1 | Port 0 | Control data | 77 | 200 |
| | | Camera data 1 | 79 | 200 |
| | | | 175 | |
| | | Camera data 2 | 89 | 200 |
| | | | 189 | |
| | | Radar data | 185 | 200 |

We deploy the parameters obtained through the AT-NFIC scheduling algorithm into the model built in OMNeT++. In addition, to verify the usability of the results more comprehensively, we similarly constructed the scenarios of frame loss and time out, using the same parameters as in Table 5. The maximum E2E latency and jitter of the streams available through simulation in OMNeT++ are shown in Figure 21.

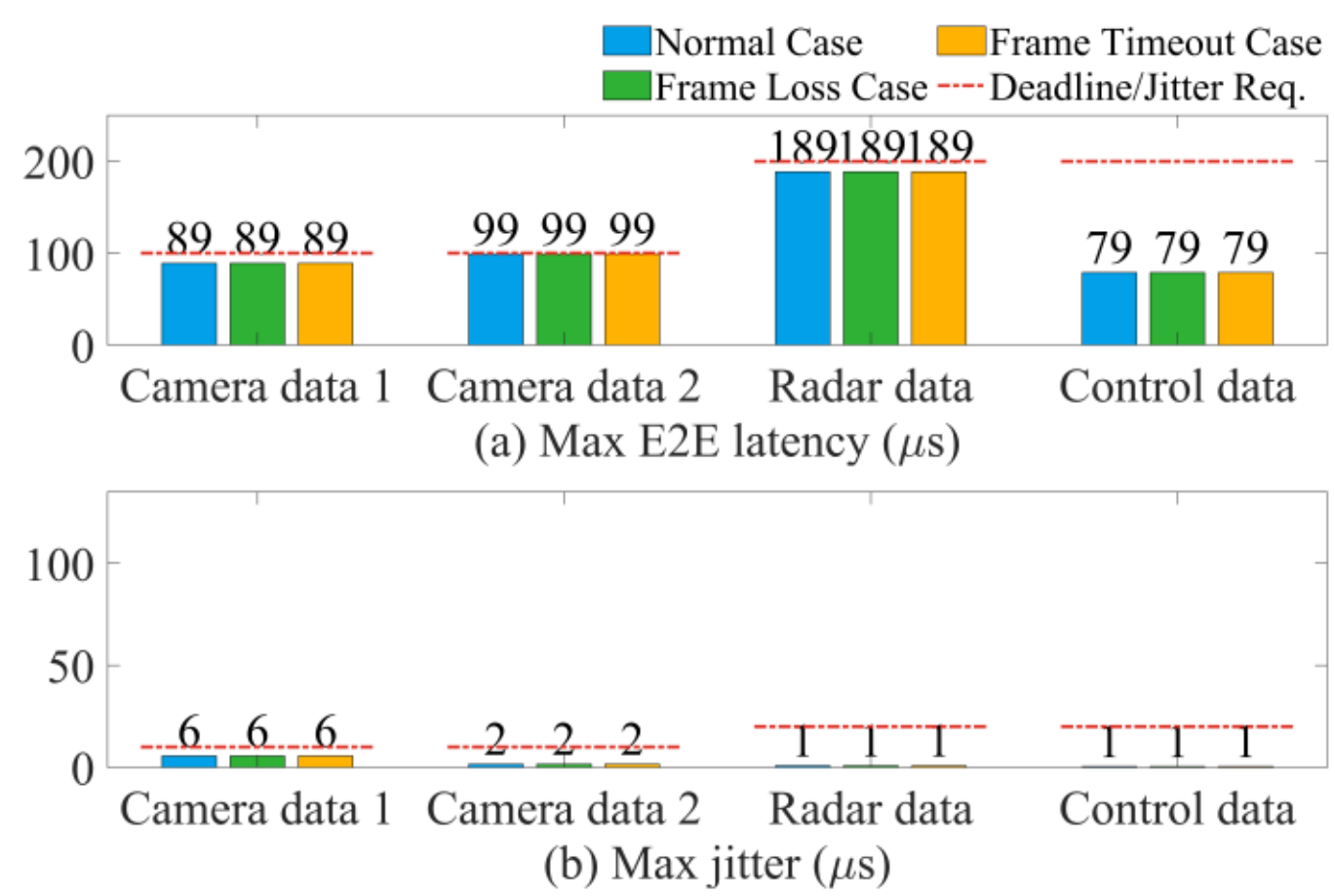

Figure 21. (a) Max E2E latency and (b) Max jitter of streams using TT-UBS, whose solution is obtained by the AT-NFIC scheduling algorithm under the normal case, frame loss case, and frame timeout case.

From Figure 21(a), we can see that even though the maximum E2E latency for Camera data 2 reaches 99us, the maximum E2E latency for the stream is within the required range, which is acceptable. Table 7 also shows that this stream is assigned a bigger *EligibilityOffset*, which can lead to this situation. The maximum E2E latency and jitter for all the streams in the normal case, the frame loss case, and the frame timeout case are within the required range, as shown in Figure 21. The above results show that the results we derived through AT-NFIC can be applied to the TT-UBS mechanism (proving the usability of the results of this scheduling algorithm).

Similar to validating the results of the AT-NFIC scheduling algorithm, we use the same methodology to validate the usability of the results of the LS-TB-NFIC scheduling algorithm. By inputting the network topology and traffic attributes into this algorithm, we obtain the parameters of the TT-UBS mechanism as shown in Table 8. Similarly, we deploy the parameters obtained through the LS-TB-NFIC scheduling algorithm into the model built in OMNeT++ and get the results as shown in Figure 22.

Table 8. The ShaperOffsetTable obtained by the LS-TB-NFIC scheduling algorithm (unit in us).

| Switch | Port | Stream | *EligibilityOffset* | *CycleTime* |
|---|---|---|---|---|
| Switch 2 | Port 0 | Control data | 2 | 200 |
| | Port 1 | Camera data 1 | 10 | 200 |
| | | | 110 | |
| | Port 2 | Camera data 2 | 20 | 200 |
| | | | 120 | |
| | Port 3 | Radar data | 4 | 200 |
| Switch 1 | Port 0 | Control data | 4 | 200 |
| | | Camera data 1 | 20 | 200 |
| | | | 120 | |
| | | Camera data 2 | 30 | 200 |
| | | | 130 | |
| | | Radar data | 8 | 200 |

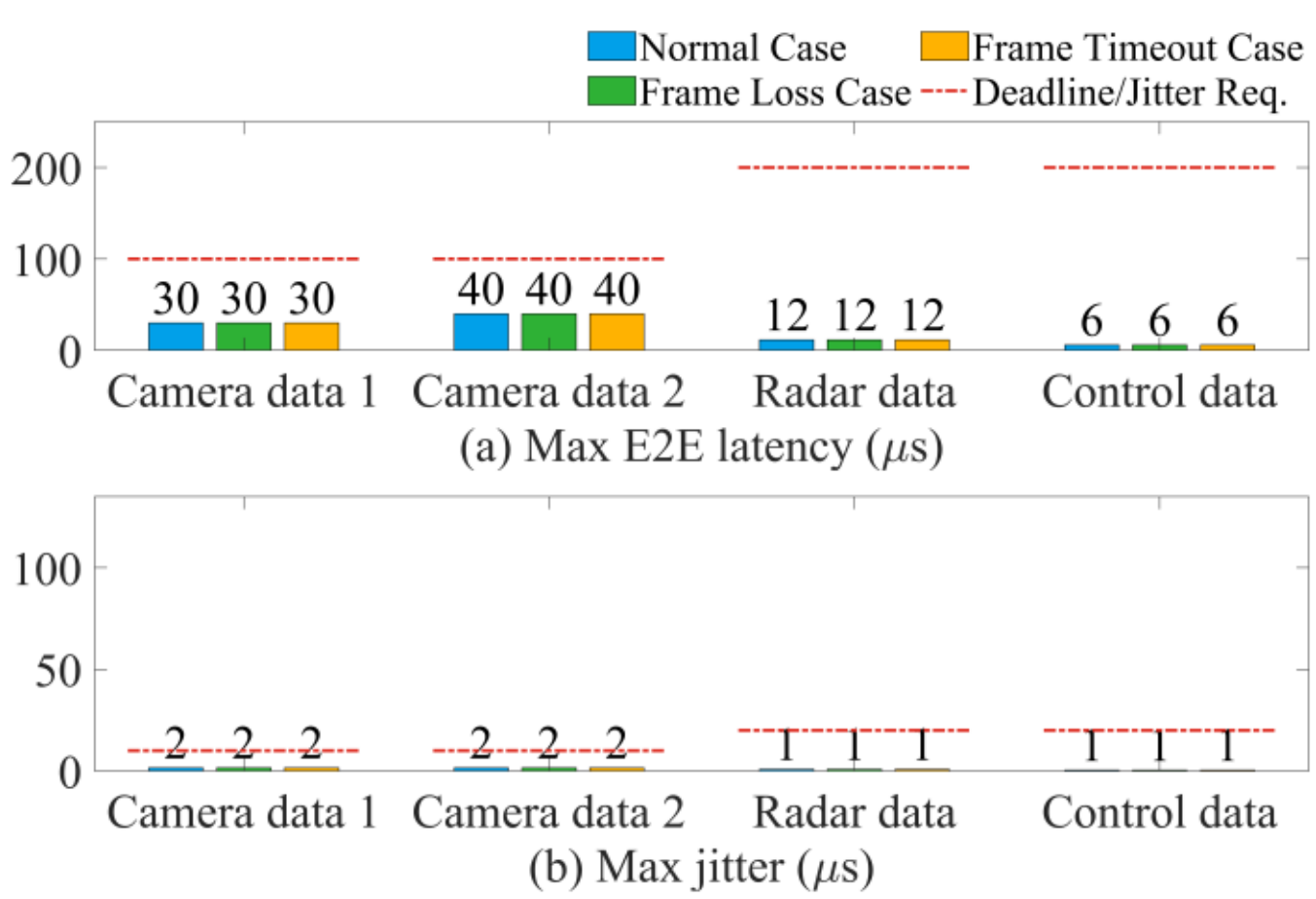

Figure 22. (a) Max E2E latency and (b) Max jitter of streams using TT-UBS, whose solution is obtained by the LS-TB-NFIC scheduling algorithm under the normal case, frame loss case, and frame timeout case.

Similarly, as shown in Figure 21, the maximum E2E latency and jitter for all streams are within the required range for the normal case, the frame loss case, and the frame timeout case. The above results can also indicate the availability of the results of the LS-TB-NFIC scheduling algorithm.

Combining with Figures 17 and 21, the averaged values of E2E latency and jitter of the streams are lower after the deployment of the results derived from the LS-TB-NFIC scheduling algorithm as compared to those derived from the SMT-WA-NFIC and AT-NFIC scheduling algorithms. This is because the first two scheduling algorithms rely on third-party solvers to produce their results. Without adding redundant constraints or objective functions, the solvers generally only produce feasible solutions that satisfy the constraints. However, algorithms without relying on third-party solvers allow more flexibility in designing the solution.

### 4.4.2 Comparison of Solving Speed

After validating the availability of results for both scheduling algorithms, we also compare their solving times with the AT and LS-TB scheduling algorithms. All experiments were run on a computer equipped with a 12th Gen Intel core i5-12400 64-bit 2.50GHz CPU with 16GB RAM. We repeatedly generate the network and solve it 20 times for each set of scenarios with the number of devices and streams. To simplify the study, the number of streams in this experiment was increased from 10 to 50, and the number of devices from 8 to 20. The solving times for AT and AT-NFIC scheduling algorithms are shown in Figure 23, and those for LS-TB and LS-TB-NFIC scheduling algorithms are in Figure 24.

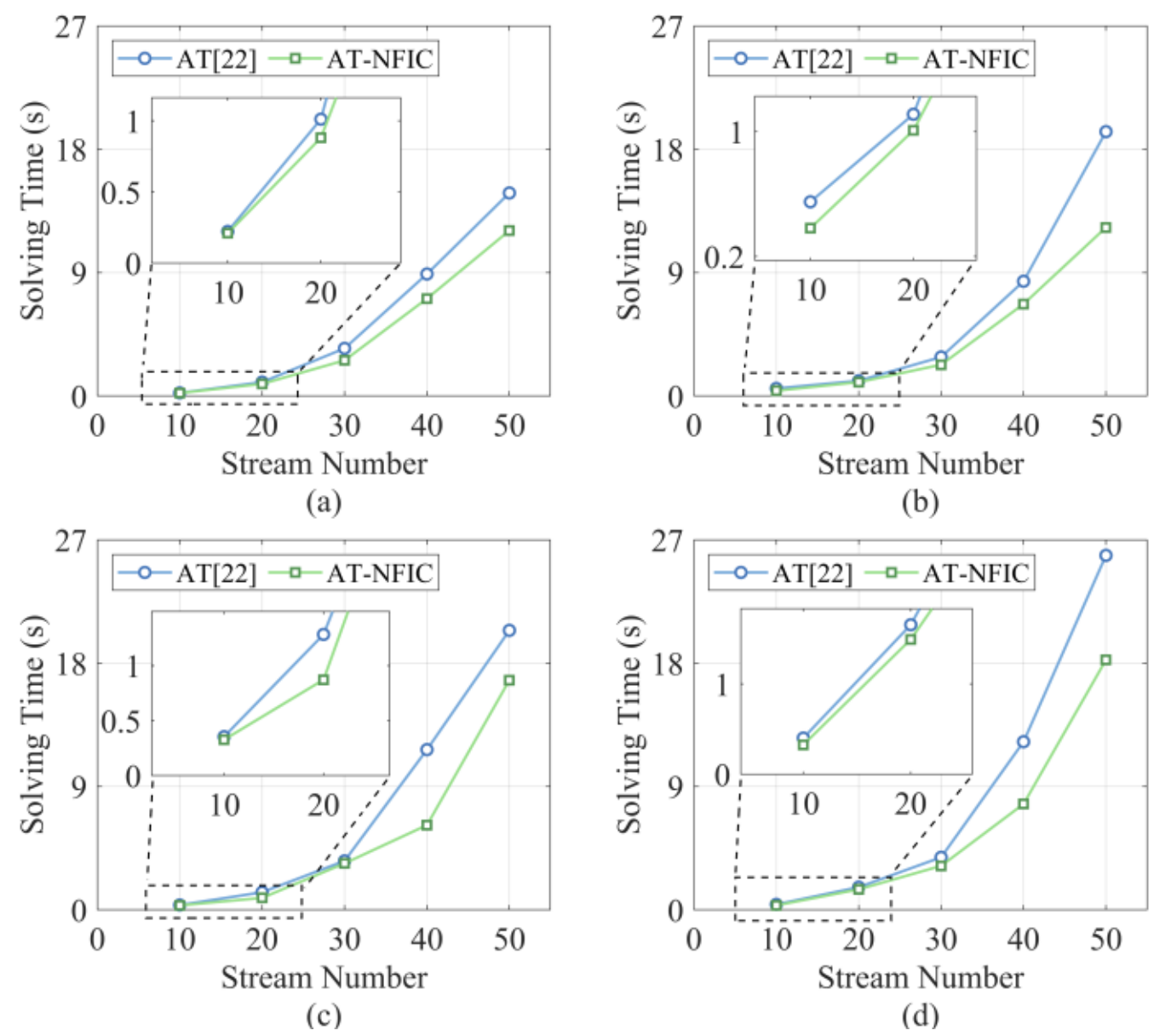

Figure 23. The variation of solving time with the increase of stream number under the AT-NFIC and AT scheduling algorithms. Device Number is (a) 8, (b) 12, (c) 16, and (d) 20.

As shown in Figure 23, the solution time of the AT and AT-NFIC algorithms increases with the number of flows, and the solution time of the AT algorithm is greater than that of the AT-NFIC algorithm. The difference between the solution times of the two algorithms is more obvious when there are more streams. In addition, it should be noted that the solution time of the AT algorithm is greater compared to the SMT-WA scheduling algorithm. For example, from Figure 20(d), when the number of devices is 20 and the number of streams is 50, the solving time of SMT-WA is within 3s. However, the solving time of AT reaches nearly 27s as shown in Figure 23(d). This indicates that the solution efficiency of the AT scheduling algorithm is lower than that of the SMT-WA, and the AT-NFIC is similar.

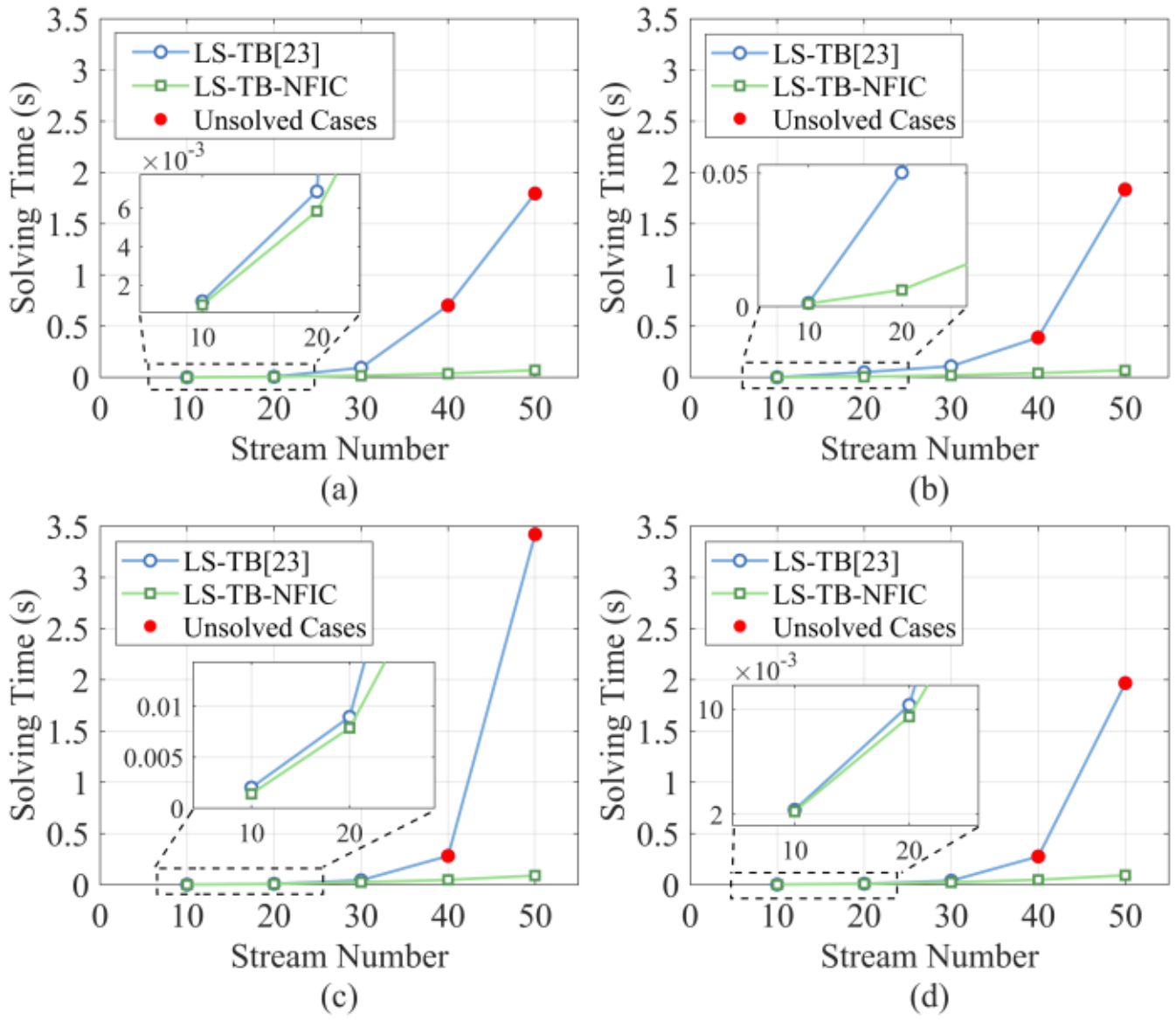

Figure 24. The variation of solving time with the increase of stream number under the LS-TB-NFIC and LS-TB scheduling algorithms. Device Number is (a) 8, (b) 12, (c) 16, and (d) 20.

Figure 24 illustrates the comparison between the LS-TB and LS-TB-NFIC algorithms in solving time. The red dots in the figure represent scenarios where one or more of the 20 experiments were unsolvable. As shown in Figure 24, after the number of streams reaches 40, the LS-TB scheduling algorithm becomes unsolvable in one or more of the 20 random scenarios. At the same time, the solving time of the AT algorithm increases significantly after the number of streams reaches 40. When the number of devices is 12 and the number of streams is 50, the solving time of the AT algorithm reaches the maximum value near 3.5s. However, even though this solution time is not high compared to the SMT-WA and AT algorithms, there may be unsolvable scenarios. In contrast, the LS-TB-NFIC algorithm has maintained a more stable and lower solving time. The average solving time of this algorithm in the experimental scenarios is no more than 0.2s, and there are no unsolvable cases. Similar to SMT-WA-NFIC, the LS-TB-NFIC algorithm also performs well in the case of network enlargement.

We summarize all the above experiments and derive the solving time for SMT-WA-NFIC, AT-NFIC, and LS-TB-NFIC scheduling algorithms, as shown in Figure 25.

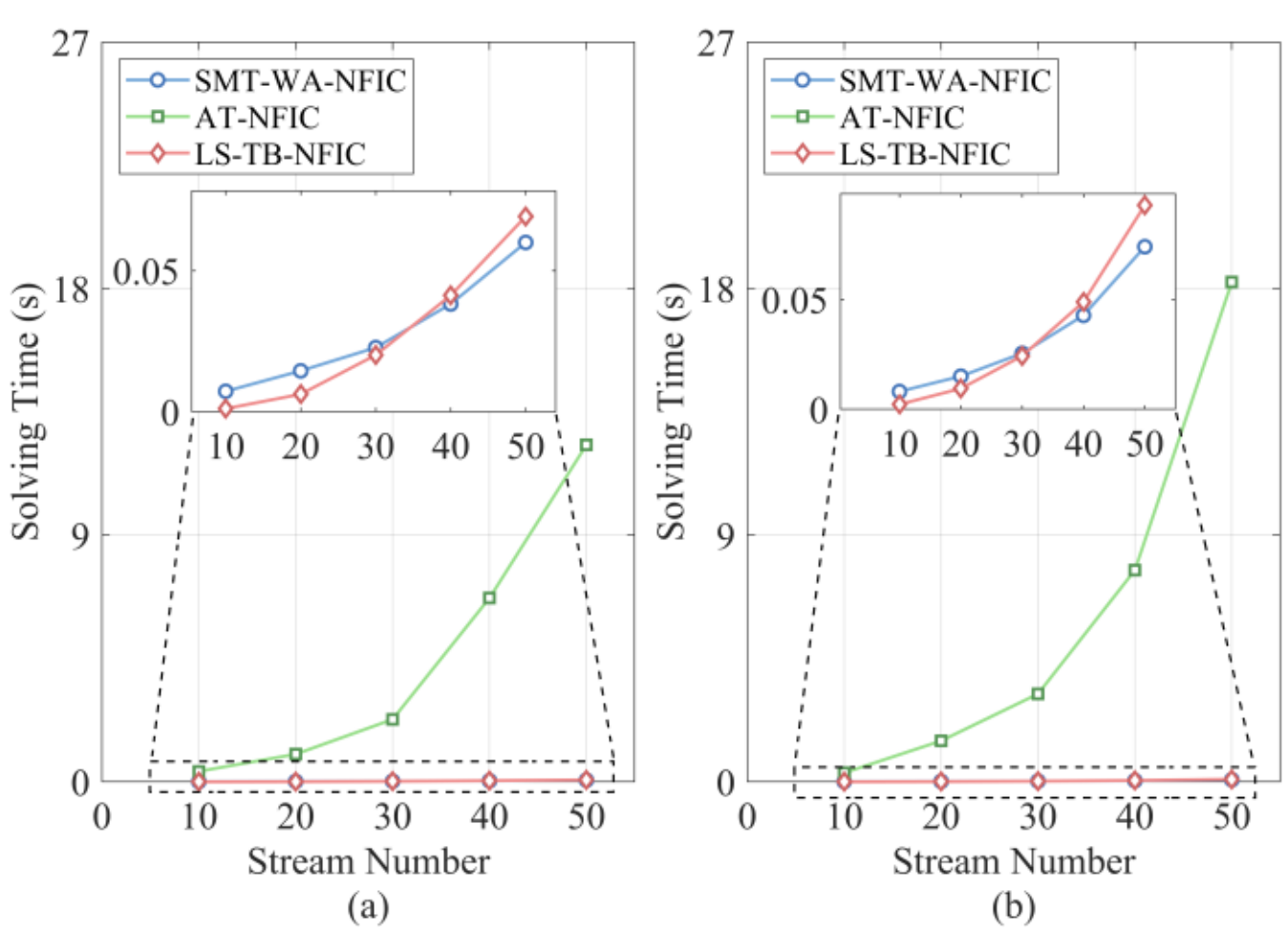

Figure 25. The variation of solving time with the increase of stream number under the SMT-WA-NFIC, AT-NFIC and LS-TB-NFIC scheduling algorithms. Device Number is (a) 12 and (b) 20.

The figure shows that the solving time of the AT-NFIC algorithm is significantly higher than the other two algorithms. In the worst case, AT-NFIC algorithm has more than 200 times the solving time of the other two algorithms. The solution times of AT-NFIC and LS-TB-NFIC algorithms are similar, and the maximum difference between them in the experiments is not more than 0.02 s. From the figure, in the experiments of this paper, the slope of the increase of the time with the number of flows of SMT-WA-NFIC is smaller than that of the LS-TB-NFIC. However, this does not conclude that the SMT-WA-NFIC algorithm is more efficient in larger networks. More experiments are needed.

## 5 Conclusion and Future Work

In this paper, we first explore how scheduling constraints change as the network grows larger by designing scenarios with increasing numbers of devices and traffic, and show that frame isolation constraints grow at a faster rate. Additionally, we illustrate the negative impact caused by TAS's inability to address timeout frames within the network.

Then, we proposed the TT-UBS mechanism to address the issues above in current TT communication. In this method, we defined the functionality of the shaper within the UBS architecture to control the transmission offset of frames through a pre-defined table. Within this approach, the shaper can discard timeout frames to prevent hindrance to the transmission of subsequent frames, thereby addressing the issues arising from frame timeouts in TAS. Then, we conducted a performance analysis of the TT-UBS mechanism, demonstrating its effectiveness in the scenarios above. We also propose the SMT-WA-NFIC scheduling algorithm to obtain the TT-UBS mechanism parameters efficiently.

Lastly, we evaluated the TT-UBS mechanism by conducting simulations in the ADAS network scenario using the modified CoRE4INET framework in OMNeT++. We compared the performance of TT-UBS and TAS in the normal case, frame loss case, and frame timeout case by modifying and establishing submodules accordingly. The results show that our proposed method performs well in normal and frame loss case and also outperforms TAS in frame timeout case. We compared the solving speed of the SMT-WA-NFIC scheduling algorithm with the SMT-WA scheduling algorithm in [20]. The results indicate that the proposed SMT-WA-NFIC algorithm has higher solving efficiency than SMT-WA. Moreover, the deployment time required for solving the TT-UBS mechanism is significantly shorter than that for TAS, and it does not exhibit a noticeable increase with network expansion. In addition, we extend the method of modifying the scheduling algorithm to two other algorithms (AT-NFIC and LS-TB-NFIC) to obtain the parameters of the TT-UBS mechanism. The comparison results show that their result can be available and they can improve the solution's efficiency, with the LS-TB-NFIC algorithm being as effective as the SMT-WA-NFIC algorithm.

In future research, we plan to conduct a timing analysis of TT-UBS to determine the upper bound of the E2E latency for each stream in the network. This analysis will provide valuable insights into the worst-case response times of the streams. To accomplish this, we will utilize a tool such as pyCPA, which is well-suited for analyzing the timing behavior of real-time systems. We will also expand the simulation scenarios to include more complex network topologies and practical application scenarios to verify the robustness and applicability of our proposed method. To further investigate the performance of our proposed method in terms of solving speed, we will conduct a comprehensive comparison between the solution method of this mechanism and other scheduling algorithms. This comparison will encompass solution speed, solution quality, solution configuration complexity, and other relevant aspects. We can also explore methods to reduce the resource consumption during the implementation process of TT-UBS and compare it with the TAS through comparative analysis. Furthermore, we aim to take the TT-UBS mechanism by implementing it on hardware platforms. This implementation will allow us to evaluate its performance and effectiveness in more complex time-critical systems, providing a realistic assessment of its capabilities.

## Acknowledgment

This work was supported by the Jiangxi Provincial Department of Science and Technology 2022 central guidance local science and technology development fund projects [grant numbers 2022ZDD03089], the Nanchang Open bidding for selecting the best candidates project [grant numbers 2023JBGS003], Jiangxi Provincial Key Research and Development Program in 2024 [grant numbers 20243BBG71010] and Jiangxi Key Laboratory of Intelligent Connected Vehicle and Powertrain System.

## References

1. Bello, L.L., Mariani, R., Mubeen, S. and Saponara, S., "Recent Advances and Trends in On-Board Embedded and Networked Automotive Systems," IEEE Transactions on Industrial Informatics, 15(2):1038-1051, 2019, doi: 10.1109/tii.2018.2879544.
2. Lo Bello, L. and Steiner, W., "A Perspective on IEEE Time-Sensitive Networking for Industrial Communication and Automation Systems," Proceedings of the IEEE, 107(6):1094-1120, 2019, doi: 10.1109/jproc.2019.2905334.
3. IEEE, "IEEE Standard for Local and Metropolitan Area Networks-Bridges and Bridged Networks," 2018.
4. Thiele, D., Ernst, R. and Diemer, J., "Formal worst-case timing analysis of Ethernet TSN's time-aware and peristaltic shapers," presented at 2015 IEEE Vehicular Networking Conference (VNC), 2015, doi: 10.1109/vnc.2015.7385584.
5. Nasrallah, A., Thyagaturu, A.S., Alharbi, Z., Wang, C., et al., "Performance Comparison of IEEE 802.1 TSN Time Aware Shaper (TAS) and Asynchronous Traffic Shaper (ATS)," IEEE Access, 7:44165-44181, 2019, doi: 10.1109/access.2019.2908613.
6. Wang, B., Luo, F. and Fang, Z., "Performance Analysis of IEEE 802.1Qch for Automotive Networks: Compared with IEEE 802.1 Qbv," presented at 2021 IEEE 4th International Conference on Computer and Communication Engineering Technology (CCET), 2021, doi: 10.1109/ccet52649.2021.9544333.
7. Maxim, D. and Song, Y.-Q., "Delay analysis of AVB traffic in time-sensitive networks (TSN)," presented at Proceedings of the 25th International Conference on Real-Time Networks and Systems, 2017, doi: 10.1145/3139258.3139283.
8. Ashjaei, M., Murselovic, L. and Mubeen, S., "Implications of Various Preemption Configurations in TSN Networks," IEEE Embedded Systems Letters, 14(1):39-42, 2022, doi: 10.1109/les.2021.3103061.
9. Luo, F., Wang, Z., Guo, Y., Wu, M., et al., "Research on Cyclic Queuing and Forwarding with Preemption in Time-Sensitive Networking," IEEE Embedded Systems Letters, 2023.
10. Luo, F., Wang, Z., Ren, Y., Wu, M., et al., "Simulative Assessments of Cyclic Queuing and Forwarding with Preemption in In-Vehicle Time-Sensitive Networking," 2024.
11. Leonardi, L., Bello, L.L. and Patti, G., "Bandwidth Partitioning for Time-Sensitive Networking Flows in Automotive Communications," IEEE Communications Letters, 25(10):3258-3261, 2021, doi: 10.1109/lcomm.2021.3103004.
12. Chen, Z., Lu, Y., Wang, H., Qin, J., et al., "Flow ordering problem for time-triggered traffic in the scheduling of Time-Sensitive Networking," IEEE Communications Letters:1-1, 2023, doi: 10.1109/lcomm.2023.3252626.
13. Xue, C., Zhang, T., Zhou, Y., Nixon, M., et al., "Real-time scheduling for 802.1 Qbv time-sensitive networking (TSN): A systematic review and experimental study," 2024 IEEE 30th Real-Time and Embedded Technology and Applications Symposium (RTAS), 2024.
14. Xu, L., Xu, Q., Tu, J., Zhang, J., et al., "Learning-based scalable scheduling and routing co-design with stream similarity partitioning for time-sensitive networking," IEEE Internet of Things Journal, 9(15):13353-13363, 2022.
15. Yang, Z., Zhao, Y., Dang, F., He, X., et al., "Caas: Enabling control-as-a-service for time-sensitive networking," IEEE INFOCOM 2023-IEEE Conference on Computer Communications, 2023.
16. Dürr, F. and Nayak, N.G., "No-wait packet scheduling for IEEE time-sensitive networks (TSN)," Proceedings of the 24th International Conference on Real-Time Networks and Systems, 2016.
17. Zhang, Y., Xu, Q., Xu, L., Chen, C., et al., "Efficient flow scheduling for industrial time-sensitive networking: A divisibility theory-based method," IEEE Transactions on Industrial Informatics, 18(12):9312-9323, 2022.
18. Zhang, Y., Xu, Q., Wang, S., Chen, Y., et al., "Scalable no-wait scheduling with flow-aware model conversion in time-sensitive networking," GLOBECOM 2022-2022 IEEE Global Communications Conference, 2022.
19. Yun, Q., Xu, Q., Zhang, Y., Chen, Y., et al., "Flexible switching architecture with virtual-queue for time-sensitive networking switches," IECON 2021–47th Annual Conference of the IEEE Industrial Electronics Society, 2021.
20. Craciunas, S.S., Oliver, R.S., Chmelík, M. and Steiner, W., "Scheduling Real-Time Communication in IEEE 802.1Qbv Time Sensitive Networks," Proc. RTNS, 2016, doi: 10.1145/2997465.2997470.
21. Vlk, M., Hanzálek, Z. and Tang, S., "Constraint programming approaches to joint routing and scheduling in time-sensitive networks," Computers & Industrial Engineering, 157:107317, 2021.
22. Serna Oliver, R., Craciunas, S.S. and Steiner, W., "IEEE 802.1Qbv Gate Control List Synthesis Using Array Theory Encoding," presented at 2018 IEEE Real-Time and Embedded Technology and Applications Symposium (RTAS), 2018, doi: 10.1109/rtas.2018.00008.
23. Vlk, M., Brejchová, K., Hanzálek, Z. and Tang, S., "Large-scale periodic scheduling in time-sensitive networks," Computers & Operations Research, 137:105512, 2022.
24. Jin, X., Xia, C., Guan, N., Xu, C., et al., "Real-time scheduling of massive data in time sensitive networks with a limited number of schedule entries," IEEE access, 8:6751-6767, 2020.
25. Bujosa, D., Ashjaei, M., Papadopoulos, A.V., Nolte, T., et al., "HERMES: Heuristic multi-queue scheduler for TSN time-triggered traffic with zero reception jitter capabilities," Proceedings of the 30th International Conference on Real-Time Networks and Systems, 2022.
26. Ergenc, D., Brulhart, C., Neumann, J., Kruger, L., et al., "On the Security of IEEE 802.1 Time-Sensitive Networking," presented at 2021 IEEE International Conference on Communications Workshops (ICC Workshops), 2021, doi: 10.1109/ICCWorkshops50388.2021.9473542.
27. Luo, F., Wang, Z. and Zhang, B., "Impact analysis and detection of time-delay attacks in time-sensitive networking," Computer Networks, 2342023, doi: 10.1016/j.comnet.2023.109936.

28. Alghamdi, W. and Schukat, M., "Precision time protocol attack strategies and their resistance to existing security extensions," Cybersecurity, 4(1):1-17, 2021, doi: 10.1186/s42400-021-00080-y.
29. Ergenç, D., Brülhart, C., Neumann, J., Krüger, L., et al., "On the security of IEEE 802.1 time-sensitive networking," 2021 IEEE International Conference on Communications Workshops (ICC Workshops), 2021.
30. Specht, J. and Samii, S., "Urgency-Based Scheduler for Time-Sensitive Switched Ethernet Networks," presented at 2016 28th Euromicro Conference on Real-Time Systems (ECRTS), 2016, doi: 10.1109/ecrts.2016.27.
31. Thiele, D. and Ernst, R., "Formal worst-case performance analysis of time-sensitive Ethernet with frame preemption," presented at 2016 IEEE 21st International Conference on Emerging Technologies and Factory Automation (ETFA), 2016, doi: 10.1109/ETFA.2016.7733740.
32. Steinbach, T., Kenfack, H.D., Korf, F. and Schmidt, T.C., "An extension of the OMNeT++ INET framework for simulating real-time ethernet with high accuracy," Proceedings of the 4th International ICST Conference on Simulation Tools and Techniques, 2011.
33. Luo, F., Wang, B., Fang, Z., Yang, Z., et al., "Security Analysis of the TSN Backbone Architecture and Anomaly Detection System Design Based on IEEE 802.1Qci," Security and Communication Networks, 2021:1-17, 2021, doi: 10.1155/2021/6902138.